\documentclass[floatfix,%
 aip,
 amsmath,amssymb,
 reprint,%
]{revtex4-1}
\pdfoutput=1
\usepackage{graphicx}
\usepackage{dcolumn}
\usepackage{bm}

\usepackage[utf8]{inputenc}
\usepackage[T1]{fontenc}
\usepackage{mathptmx}
\usepackage{etoolbox}

\makeatletter
\def\@email#1#2{%
 \endgroup
 \patchcmd{\titleblock@produce}
  {\frontmatter@RRAPformat}
  {\frontmatter@RRAPformat{\produce@RRAP{*#1\href{mailto:#2}{#2}}}\frontmatter@RRAPformat}
  {}{}
}%
\makeatother
\begin{document}

\preprint{AIP/123-QED}

\title{The Development of a Multi-Physics Approach for Modelling the Response of Aerospace Fastener Assemblies to Lightning Attachment}

\author{William Paul Bennett} \thanks{Corresponding Author; wpb22@cam.ac.uk}
\author{Stephen Timothy Millmore}
\author{Nikolaos Nikiforakis}
\affiliation{University of Cambridge,\\ Laboratory for Scientific
  Computing, Cavendish Laboratory, Department of Physics,\\ Cambridge, CB3 0HE, UK }


\begin{abstract}

\noindent This work is concerned with the development of a numerical modelling approach for studying the time-accurate response of aerospace fasteners subjected to high electrical current loading from a simulated lightning strike. The electromagnetic, thermal and elastoplastic response of individual fastener components is captured by this method allowing a critical analysis of fastener design and material layering. Under high electrical current loading, ionisation of gas filled cavities in the fastener assembly can lead to viable current paths across internal voids. This ionisation can lead to localised pockets of high pressure plasma through the Joule heating effect. The multi-physics approach developed in this paper extends an existing methodology that allows a two-way dynamic non-linear coupling of the plasma arc, the titanium aerospace fastener components, the surrounding aircraft panels, the internal supporting structure and internal plasma-filled cavities. Results from this model are compared with experimental measurements of a titanium fastener holding together carbon composite panels separated by thin dielectric layers. The current distribution measurements are shown to be accurately reproduced. A parameter study is used to assess the internal cavity modelling strategy and to quantify the relation between the internal cavity plasma pressure, the electrical current distribution and changes in the internal cavity geometry.


\end{abstract}

\maketitle

\section{Introduction}

Composite materials now make up more than 50\% of a modern aircraft
design by weight~\cite{krauklis2020composite}. This is due to higher specific strength 
and better fatigue properties for high
tension components than conventional aluminium alternatives. Composite materials also have lower electrical and thermal conductivity than the
supplanted aluminium, which can adversely affect the response of
aircraft components to lightning strikes. The initial stages of a
lightning strike results in a large current flow through the aircraft
skin, which, for materials with low electrical conductivity can result
in large energy input through Joule heating. In Carbon Fibre
Reinforced Polymer (CFRP), for example, the high energy input can
result in fibre fracture, resin decomposition, delamination and
thermal ablation.

The interaction of lightning with aircraft exterior surfaces can be
further complicated by the integration of conductive components, such
as metallic fasteners, with {higher} conductivity than
the surrounding composite substrates. Aerospace fasteners, used to
join skin panels, are commonly manufactured {from}
titanium alloys that are lightweight, strong and corrosion
resistant. These can, however, act as a preferred pathway for the
current to access the internal airframe and embedded carbon composite
panels. High current flow through a fastener can arise either from a
direct attachment of a {lightning} arc to the fastener
head or indirectly as current conducted from a remote attachment
site. In addition to paint, panel and sealant damage, the high current
flow in the metallic fastener can cause a thermal ejection of hot
particles (energetic discharge) from the {interfaces}
between fastener components. Energetic discharge from fastener
assemblies can represent a potential threat in safety critical regions
of an aircraft, such as in integrated fuel tanks where significant
fuel vapour is present.

Chemartin {et al.}~\cite{chemartin2012direct} outline three important
mechanisms through which a fastener can undergo energetic
discharge. The first, termed `outgassing', results from the current
passage across small resistive gaps between the fastener and
skin. The formation of a plasma in the gap, and
subsequent Joule heating, increases the internal plasma pressure
until a blow out of sparks and hot gas occurs at the component
interface. Further information regarding the characteristics and
causes of outgassing (also known as pressure sparking) is provided in
the comprehensive review of
Evans~\cite{evans2018thesisCharacterisation}.  The second energetic
discharge mechanism outlined in Chemartin {et al.}~\cite{chemartin2012direct} is thermal sparking between contacting components,
i.e., the fastener nut and rib. Odam et al.~\cite{odam1991lightning,
  odam2001factors} suggests that thermal sparking occur when a very high
current is forced to cross a joint between two conducting materials,
which have imperfect mating between their surfaces.  This process is
noted to be distinct from voltage sparking, which occurs when the
voltage difference between two conducting materials is sufficient to
break down the intervening medium, whether this is air or another
dielectric medium.  The final energetic discharge mechanism outlined in Chemartin {et al.}~\cite{chemartin2012direct} is edge
glow. This is defined by Revel et al.~\cite{revel2016edge} as
consisting of a bright glow combined with strong material ejections,
and occurs on the edges of composite materials. Two
mechanisms responsible for the presence of edge glow are proposed in the available literature. The first of
these occurs when the potential difference between adjacent composite
plies exceeds a threshold value, irrespective of whether a
pre-existing contact between the plies is present. The second
mechanism occurs when the power deposition into the substructure is
above a threshold power value that produces a glow due to heating of
sealant.

These energetic discharge mechanisms can been distinguished using an appropriate experimental approach. 
Day and Kwon~\cite{kwon2009optical} describe a method which analyses light
over a narrow spectral range using a spectrometer and can identify hot
particle ejection, arcing and edge glow. Work by Haigh et
al.~\cite{haigh1991measurements}, in contrast, applies two-colour
spectroscopy to estimate the temperature of sparks emitted using
red/blue ratios, comparing with baseline nickel and tungsten
sparks. Later work by Haigh et al.~\cite{haighCrookTerzino2013} uses
photography to detect light sources that are potential ignition
hazards on a T-joint section with multiple fasteners. Focusing
specifically on outgassing events, Mulazimoglu and
Haylock~\cite{mulazimoglu2011recent} relate sparking intensity to the
fastener material and geometry choice using energy dispersion
spectrometer chemical analysis, and determine that the principle
constituent of the ejected particle debris in question is polysulphide sealant,
with small quantities of metallic droplets and carbon fibre
particles. They surmise that the chemical composition of the debris
mean that electrical arcing occurs between the bolt and the CFRP hole
surface. The ablated material is then carried by hot gases during the
outgassing ejection event due to the arcing pressure. The
microstructure of the resulting damage is analysed using scanning
electron microscopy and the outgassing characteristics that result
from deliberate design additions are analysed. These additions include
the introduction of metallic sleeve components, dielectric bolt head
coverings and bolt-line metallic meshes.

The wide range of potential fastener configurations, along with the
various mechanisms through which sparking can occur, mean that
computational simulation can provide an efficient and cost effective
technique for rapidly exploring the available parameter space.
Computational techniques can also provide a useful tool in the design
of experimental testing for proposed fastener configurations. Finite
element methods, for example, are a common, single-material approach to model the
effects of high current flow through carbon composite substrates.
This is achieved through prescribing a current waveform along the
upper surface, see, for example Ogasawara et
al.~\cite{ogasawara2010coupled}, Abdelal and
Murphy~\cite{abdelal2014nonlinear}, Guo et
al.~\cite{guo2017comparison}, Dong et al.~\cite{dong2015coupled}, Wang
et al.~\cite{wang2014ablation} and Liu et al.~\cite{liu2015combining}
and for commercial software by Wang and
Pasiliao~\cite{wang2018modeling}, Kamiyama et
al.~\cite{kamiyama2018delamination, kamiyama2019damage}, Fu and
Ye~\cite{fu2019modelling} and Evans et
al.~\cite{evans2014comsolansys}. The prescription of a current
waveform along the upper surface of a composite material is perhaps
most suited to cases in which the damage to individual ply and resin
layers is of direct interest, since inter-ply loading and damage
characteristics can be efficiently modelled using modest computational
resources for comparison with experimental results. Modelling a
lightning strike using this approach in isolation can, however,
neglect the dynamic change in current and pressure loading on the
upper surface of the substrate by an evolving plasma arc, and the
non-linear feedback from these changes, which in turn affect the arc
behaviour.

To allow the evolution of the plasma arc to effect the time-dependent
substrate current and pressure loading, the `co-simulation' approach
couples two software packages or approaches. Using this technique a
magnetohydrodynamic (MHD) code can be used to describe the evolution
of temperature, pressure and current density within the arc. The
results of running the MHD code are then fed as initial and boundary
conditions to a second simulation that models the mechanical, thermal
and electrodynamic evolution of the substrate. Examples of this
approach include Tholin et al.~\cite{tholin2015numerical}, who couple
two distinct software packages (C\`edre and Code--Saturn) to model the
plasma arc attachment to single material substrates, and Millen et
al.~\cite{millen2019sequential} who couple two commercial software
packages (COMSOL Multiphysics and Abaqus FEA) to model the mechanical
loading and electromagnetic effects on a carbon composite
substrate. Kirchdoerfer et al.~\cite{kirchdoerfer2017,
  kirchdoerfer2018cth} apply the co-simulation approach to aerospace
fasteners, coupling the results from COMSOL Multiphysics with a
research shock-physics code developed at Sandia National Laboratories
(CTH). The electric and magnetic fields, and current density, are
solved in COMSOL and used to determine Joule heating effects. One-way
coupling is then applied with the CTH solver using an effective
heating power, computed from the modelled Joule heating, allowing
the simulation of the fluid-structure interaction. This one-way
coupled solution is used to determine the temperature and pressure
rise in an internal cavity between a bolt, nut, and surrounding CFRP
panels, with the final pressure rise being compared to experimental
results.

The co-simulation approach results in a one-directional coupling
between materials where the substrate behaviour does not influence the
evolution of the arc. However, experimental results, such as the
optical measurements of Martins~\cite{martins2016etude}, indicate that
changes in the electrical conductivity and substrate shape can alter
the arc attachment characteristics. This work uses a multi-physics
methodology introduced in Millmore and
Nikiforakis~\cite{millmore2019multi}, to simulate a dynamic
non-linearly coupled system comprising the plasma arc, titanium
aerospace fastener components, surrounding aircraft panels and the
internal supporting structure. The electromagnetic, thermal and
elastoplastic response of individual fastener components is captured
by this method, allowing a critical analysis of fastener design and
material layering. Dynamic feedback between the components is achieved
in this multi-physics approach by simultaneously solving the governing
hyperbolic partial differential equations for each material in a
single system. The non-linear dynamic feedback between adjacent
materials achieved by this approach provides a distinct improvement
over existing numerical methods for modelling lightning strike
attachment. The underlying numerical methods and implementation used
in this paper are outlined in Millmore \&
Nikiforakis~\cite{millmore2019multi}, and extended in Michael et
al.~\cite{michael2019multi} and Tr{\"a}uble et
al.~\cite{trauble2021improved}.  Millmore and
Nikiforakis compare numerical results from
the non-linear multi-physics approach used in this paper with the
optical measurements of Martins~\cite{martins2016etude}, for a plasma
arc attachment to a single material substrate, and demonstrate that
the two-way interaction between the substrate and plasma is accurately
captured by this method.

The key aim of this work is to use this approach to model the rise in pressure within an internal cavity between a titanium
fastener and a CFRP panel. The breakdown of air in this cavity
requires a strategy for defining an internal plasma region, and the
influence of parametric changes in the cavity geometry on the pressure
rise through Joule heating can be studied. This mechanism is
acknowledged by Chemartin et al.~\cite{chemartin2012direct} and
Evans~\cite{evans2018thesisCharacterisation} as a major contributing
factor in outgassing, hence this paper focuses on this mechanism
over thermal sparking and edge glow.  An overview of the multi-physics
methodology is given in section~\ref{sec:MathematicalModel} and an
assessment of the methodology for modelling lightning strikes on
aerospace fasteners is made in
section~\ref{sec:FastenerModelling}. The model is validated by
comparing to experimental measurements of a fastener holding together
carbon composite and aluminium panels, electrically isolated from each
other by a dielectric layer. The multi-physics methodology is then
exercised in section~\ref{sec:SensitivityStudies} to investigate how
parametric changes in the design of an idealised fastener influence
the pressure loading, component temperature rise and electrical
current path characteristics. This study considers a variety of
fastener design and material layering choices, and permits the
pressure rise in confined internal plasma regions to be numerically
quantified. Section~\ref{sec:Conclusions} provides a summary and an
outlook to future work.


\section{Mathematical Model Description}
\label{sec:MathematicalModel}

This section gives an overview of the mathematical model used to
simulate the response of aerospace fastener assemblies to lightning attachment.

\subsection{{ Plasma model}}
\label{sec:PlasmaModelling}

The plasma arc is described through a system of equations suitable for
simulating the evolution and ionisation of air caused by an electric
discharge. This model must describe the change in the chemical
composition of the plasma under the high temperatures of the arc.
Additionally electromagnetic effects can have a strong influence on
the arc evolution. This requires an MHD formulation which includes the
effects of current flow in the arc through the Lorentz and Joule
effects.  For lightning plasma, this system can be assumed to be under
local thermodynamic
equilibrium~\cite{tholin2015numerical,martins2016etude}. The governing
system of equations is therefore given by

 \begin{equation}
\frac{\partial \rho}{\partial t} + \nabla \cdot \left(\rho{\bf v}\right) = 0, 
\label{eq:mhdrho}
\end{equation}
 \begin{equation}
\frac{\partial }{\partial t}\left(\rho {\bf v}\right) + \nabla\cdot\left(\rho{\bf v}\otimes{\bf v}\right) + \nabla p = {\bf J}\times{\bf B},
\label{eq:mhdcont}
\end{equation}
 \begin{equation}
\frac{\partial E}{\partial t} + \nabla \cdot \left[\left(E + p\right) {\bf v}\right] = {\bf v} \cdot \left({\bf J} \times {\bf B}\right) + \eta {\bf J} \cdot {\bf J} - S_r,
\label{eq:mhdene}
\end{equation}
 \begin{equation}
-\nabla^2 {\bf A} = \mu_0 {\bf J}.
\label{eq:mhdmag}
\end{equation}

where $\rho$ is density, ${\bf v}$ is velocity, $p$
is the pressure, $E$ is the total energy, ${\bf J}$
is the current density, ${\bf B}$ is the magnetic field, $\eta$ is the
resistivity of the plasma, $S_r$ is a term for the radiative losses
from a heated material, and ${\bf A}$ is the magnetic vector
potential, related to the magnetic field through
${\bf B} = \nabla \times {\bf A}$. The current density is governed by
the continuity equation.
\begin{equation}
\nabla \cdot {\bf J} = -\nabla \cdot \left(\sigma \nabla \phi\right) = 0
\label{eq:current-cont}
\end{equation}

where $\phi$ is the electric potential and $\sigma = 1/\eta$ is the
electrical conductivity of the plasma arc. 

\subsubsection{Equation of state}
\label{sec:equation-state}

The system of equations~(\ref{eq:mhdrho})--(\ref{eq:mhdmag}) comprises
8 equations for 10 unknown variables, $\rho$, $E$,
$p$ and the vectors ${\bf v}${, ${\bf B}$ and $\phi$}.
These are closed using equation~(\ref{eq:current-cont}) and an
equation of state which describes the thermodynamic properties of the
system. The equation of state is typically written in the form
$p = p\left(e, \rho\right)$, where $e$ is the specific internal
energy, and is related to the total energy through
$E = \rho e + \frac{1}{2}\rho v^2$.  The equation of state of a plasma
is complex, since its thermodynamic properties depend strongly on the
degree of ionisation.

To capture this behaviour, a tabulated equation of state for air
plasma is used in this paper.  This was developed by Tr{\"a}uble et
al.~\cite{trauble2021improved}, based on the theoretical model of
d'Angola et al.~\cite{d2008thermodynamic}. This considers the 19 most
important species present in an air plasma at temperatures up to
$60,000$~K over a pressure range of $0.01<p<100$~atm.


\subsection{Elastoplastic model}
\label{sec:MultimaterialModelling}

In this work, the material substrate uses an elastoplastic solid
model described by Barton et al.~\cite{Barton20097046} and
Schoch et al.~\cite{schoch2013eulerian, schoch2013propagation}, based
on the formulation by Godunov and
Romenskii~\cite{godunov1972nonstationary}. The plasticity model
follows the work of Miller and Colella~\cite{miller2001high}. The
elastoplastic implementation used in the present work is described in
Michael et al.~\cite{michael2019multi}, therefore only a brief
overview is provided here for completeness.

To account for the material deformation, the elastic deformation
gradient is defined as
\begin{equation}
  \label{eq:elas-def-grad}
  \mathrm{F}^e_{ij} = \frac{\partial x_i}{\partial X_j}
\end{equation}
which maps a coordinate in the original configuration, $X_i$, to its
evolved coordinate in the deformed configuration, $x_i$.  The
deformation gradient, along with the momentum, energy and a scalar material
history parameter, $\kappa$, give a hyperbolic system of conservation
laws
 \begin{equation}
  \frac{\partial \rho u_i}{\partial t} + \frac{\partial}{\partial
    x_k}\left(\rho u_iu_k - \sigma_{ik}\right) = 0,
  \label{eq:solid-form-1}
\end{equation}
\begin{equation}
  \frac{\partial \rho E}{\partial t} + \frac{\partial}{\partial
    x_k}\left(\rho E u_k - u_i\sigma_{ik}\right) = \eta J_i J_i
  \label{eq:solid-form-2}
\end{equation}
\begin{equation}
  \frac{\partial \rho \mathrm{F}_{ij}^e}{\partial t} +
  \frac{\partial}{\partial x_k}\left(\rho u_k\mathrm{F}_{ij}^e - \rho
    u_i\mathrm{F}_{kj}^e\right)
  = -u_i\frac{\partial \rho \mathrm{F}_{kj}^e}{\partial x_k} + \mathrm{P}_{ij},
  \label{eq:solid-form-2_5}
 \end{equation}
\begin{equation}
  \frac{\partial \rho \kappa}{\partial t} + \frac{\partial}{\partial
    x_i} \left( \rho u_i \kappa \right) = \rho \dot{\kappa}.
  \label{eq:solid-form-3}
\end{equation}
\begin{equation}
  -\nabla^2  A_i = \mu_0 J_i
  \label{eq:solid-form-4}
\end{equation}
where $\sigma$ is the stress tensor, given by
\begin{equation}
  \label{eq:stress-defn}
  \sigma_{ij} = \rho \mathrm{F}^e_{ik} \frac{\partial e}{\partial \mathrm{F}^e_{jk}}
\end{equation}
and $e$ is the specific internal energy.  The scalar material history,
$\kappa$, tracks work hardening of the material through plastic
deformation.  Source terms associated with the plastic update are
denoted $\mathrm{\bf P}$.  The density is given by
\begin{equation}
  \label{eq:solid-density}
  \rho = \frac{\rho_0}{\mathrm{det}\,{\mathrm{\bf F}}^e}
\end{equation}
where $\rho_0$ is the density of the initial, unstressed material and
the system is coupled with compatibility constraints on the
deformation gradient
\begin{equation}
  \label{eq:compat-const}
  \frac{\partial \rho \mathrm{F}_{ji}}{\partial x_j} = 0
\end{equation}
which ensure deformations of the solid remain physical.


\subsubsection{Equations of state for elastoplastic materials}
\label{sec:SubstrateElectrodynamics}

As with the plasma model, in order to describe the thermodynamic
properties of the model, and to close the system of
equations~(\ref{eq:solid-form-1})--(\ref{eq:solid-form-4}), an
equation of state is required.  A variety of solid materials are used
in aerospace fasteners, in particular, aluminium, composite materials,
titanium, dielectric coatings and sealants.  Additionally,
experimental studies include an electrode, which is typically tungsten
or steel. This is used to generate a plasma arc and evolution
within this electrode does not affect the simulation dynamics (and
damage issues are not an issue for simulation purposes).  Therefore
any conductive metal is typically used as a substitute.

Aluminium is typically described through a Romenskii equation of
state~\cite{Barton20105518}, for which the specific internal energy of
the metal is given by
\begin{equation}
\label{eq:romenskii-eos}
\begin{array}{c}
  e = \frac{K_0}{2\alpha^2}\left({\mathcal{I}_3^{\alpha/2} - 1}\right)^2 + c_v T_0
  \mathcal{I}_3^{\gamma/2} \left({\mathrm{exp}\left({S/c_v}\right)-1}\right) + \\
  \frac{B_0}{2}\mathcal{I}_3^{\left({\beta/2}\right)}{\frac{\mathcal{I}_1^2}{3}-\mathcal{I}_2} 
  \end{array}
\end{equation}
where
\begin{equation}
  \label{eq:romesnk-coeffs}
  K_0 = c_0^2 - \frac{4}{3}b_0^2, \qquad B_0 = b_0^2
\end{equation}
and these are the bulk speed of sound and the shear wave speed
respectively, with $c_0$ being the sound speed, $S$ is the entropy,
$c_v$ is the heat capacity at constant volume and $T_0$ a reference
temperature.  The constants $\alpha$, $\beta$ and $\gamma$ are related
to the non-linear dependence of sound speeds on temperature and
density, and must be determined experimentally for each material.  The
quantities $\mathcal{I}_K$ are invariants of the Finger strain tensor
$G = F^{-T} F^{-1}$, and are given by
\begin{equation}
  \label{eq:finger-invar}
  \mathcal{I}_1 = \mathrm{tr}\left({G}\right), \quad \mathcal{I}_2 =
  \frac{1}{2}\left[{\left({\mathrm{tr}\left({G}\right)}\right)^2 - tr\left({G^2}\right)}\right],
  \quad \mathcal{I}_3 = \mathrm{det}\left|{G}\right|.
\end{equation}
The entropy is computed from the primitive variables and a reference
density $\rho_0$,
\begin{equation}
  \label{eq:romensk-entropy}
  S = c_v \mathrm{log} \left({\frac{\frac{p}{\rho} - K_0 \alpha
    \left({\frac{\rho}{\rho_0}}\right)^\alpha
    \left[{\left({\left({\rho}\right){\rho_0}}\right)^\alpha - 1}\right]}{\gamma c_v T_0
    \left({\frac{\rho}{\rho_0}}\right)^\gamma} + 1}\right).
\end{equation}
The parameters for aluminium, as used in Barton et al.~\cite{barton2010eulerian}, are given by
\begin{equation}
  \label{eq:al1100-params}
  \begin{array}{c}
  \rho_0=2710 \text{\,kg\,m}^{-3},\,
  c_v=900\text{\,J\,kg$^{-1}$\,K$^{-1}$} \\
  T_0=300\text{\,K},\,b_0=3160\text{\,m\,s}^{-1} \\ 
  c_0=6220\text{\,m\,s}^{-1},\, \alpha=1 \\ 
  \beta=3.577,\, \gamma=2.088.
  \end{array}
\end{equation}

Carbon composites are anisotropic materials, and thus have a more
complex equation of state. These are described in the work of
Lukyanov~\cite{lukyanov2010equation}, though their implementation in
this model is, at present, beyond the scope of this work.  Additional
work is required within the elastoplastic model described above to
deal with material anisotropy.  Following Millmore and
Nikiforakis~\cite{millmore2019multi}, an isotropic approximation to
CFRP can be made, which is suitable for modelling `with weave' and
`against weave' configurations due to the symmetries of the problem.
This isotropic approximation uses the equation of state as for
aluminium, but with electrical conductivity values that approximate
CFRP.

A Romenskii equation of state is used for modelling titanium
components due to the lack of readily available equations of state
for this material.  In the configurations considered in this work,
electromagnetic effects dominate, and thus this approximation does not
have a significant effect on the evolution.  The parameters for this
equation of state are
\begin{equation}
  \label{eq:titanium-params}
  \begin{array}{c}
  \rho_0=8030 \text{\,kg\,m}^{-3},\,
  c_v=500\text{\,J\,kg$^{-1}$\,K$^{-1}$} \\
  T_0=300\text{\,K},\, b_0=3100\text{\,m\,s}^{-1} \\
  c_0=5680\text{\,m\,s}^{-1},\, \alpha=0.596 \\ 
  \beta=2.437,\, \gamma=1.563.
  \end{array}
\end{equation}

The dielectric coatings on aircraft skins are similarly complex, often
comprising several layers of material and equations of state for these
materials are not openly available.  For such coatings used in this
work, the plastic polymethyl methylacetate (PMMA) is used, which has
been widely studied due to its use in improvised
explosives~\cite{Christou201248,hamada2004performance}.  For PMMA, a
Mie-Gr{\"u}neisen equation of state is employed which directly
relates pressure and density through
\begin{equation}
  \label{eq:mie-grun-eos}
    p = \frac{\rho_0 c_0^2}{s\left({1-s\eta}\right)}\left({\frac{1}{1-s\eta} -
    1}\right), \qquad \eta = 1 - \frac{\rho_0}{\rho}
\end{equation}
where $c_0$ and $\rho_0$ again refer to the speed of sound in the
material and the reference density, whilst $s$ is a single
experimentally determined coefficient.  These quantities are
given by
\begin{equation}
  \label{eq:pmma-mg-values}
  \rho_0 = 1180\ \mathrm{kgm}^{-3}, \quad c_0 = 2260\
  \mathrm{ms}^{-1}, \quad s = 1.82.
\end{equation}

For all materials, an electrical conductivity is also required for use
in equation~(\ref{eq:solid-form-2}).  Over the temperature ranges
observed within the substrate, this can be considered constant for all
three materials used in this work.  The electrical
conductivity employed for aluminium is
$\sigma = 3.2\times10^7$~Sm$^{-1}$ (e.g., Tholin et al.~\cite{tholin2015numerical}), for
titanium: $\sigma = 2.38\times 10^6$~Sm$^{-1}$ and for PMMA:
$\sigma = 2.6\times10^{-5}$~Sm$^{-1}$. The electrical conductivity for PMMA is several orders of magnitude lower than all other materials used and approximates a dielectric layer. The carbon
composite is modelled using the isotropic approximation of Millmore
and Nikiforakis~\cite{millmore2019multi} where, unless otherwise
stated, a bulk electrical conductivity of $4.1\times 10^4$~Sm$^{-1}$
is used. 


\subsection{Numerical approach}
\label{sec:NumericalMethod}

The coupled multi-physics system requires the solution for the plasma,
equations~(\ref{eq:mhdcont})-(\ref{eq:mhdene}), and for the
elastoplastic solid components,
equations~(\ref{eq:solid-form-1})-(\ref{eq:solid-form-3}), each of
which is a hyperbolic system of partial differential equations. In addition, we require the elliptic magnetic field, equations~(\ref{eq:mhdmag}) and
(\ref{eq:current-cont}), which exist for all materials.  The
hyperbolic equations are solved using a finite volume methodology; in
this case, a second order, slope limited centred method is employed
for solving the discrete form of the equations and updating each
material independently.

Information is passed across material boundaries using a ghost fluid
method, with level set methods (one for each material) being used to
track the location of the interfaces between each material. Dynamic
boundary conditions are applied at the interfaces using the Riemann
ghost fluid method of Sambasivan and
Udaykumar~\cite{sambasivan2009ghost1}, which provides these conditions
through a mixed-material Riemann solver. The projection method of
Lasasso et al.~\cite{losasso2006multiple} is used to correct the level
set for non-unique level-set values caused by potential numerical
approximation errors. A comprehensive overview of the multi-material
approach used in this work is given by Michael et
al.~\cite{michael2019multi}.

The elliptic equations are solved by coupling the system to a finite
element solver, in this case, the DOLFIN package for the
FEniCS framework~\cite{langtangen2016solving} is used.


\section{Experimental comparison of the numerical model}
\label{sec:FastenerModelling}

\begin{figure}[!ht]
  \centering
   \includegraphics[width=0.5\textwidth]{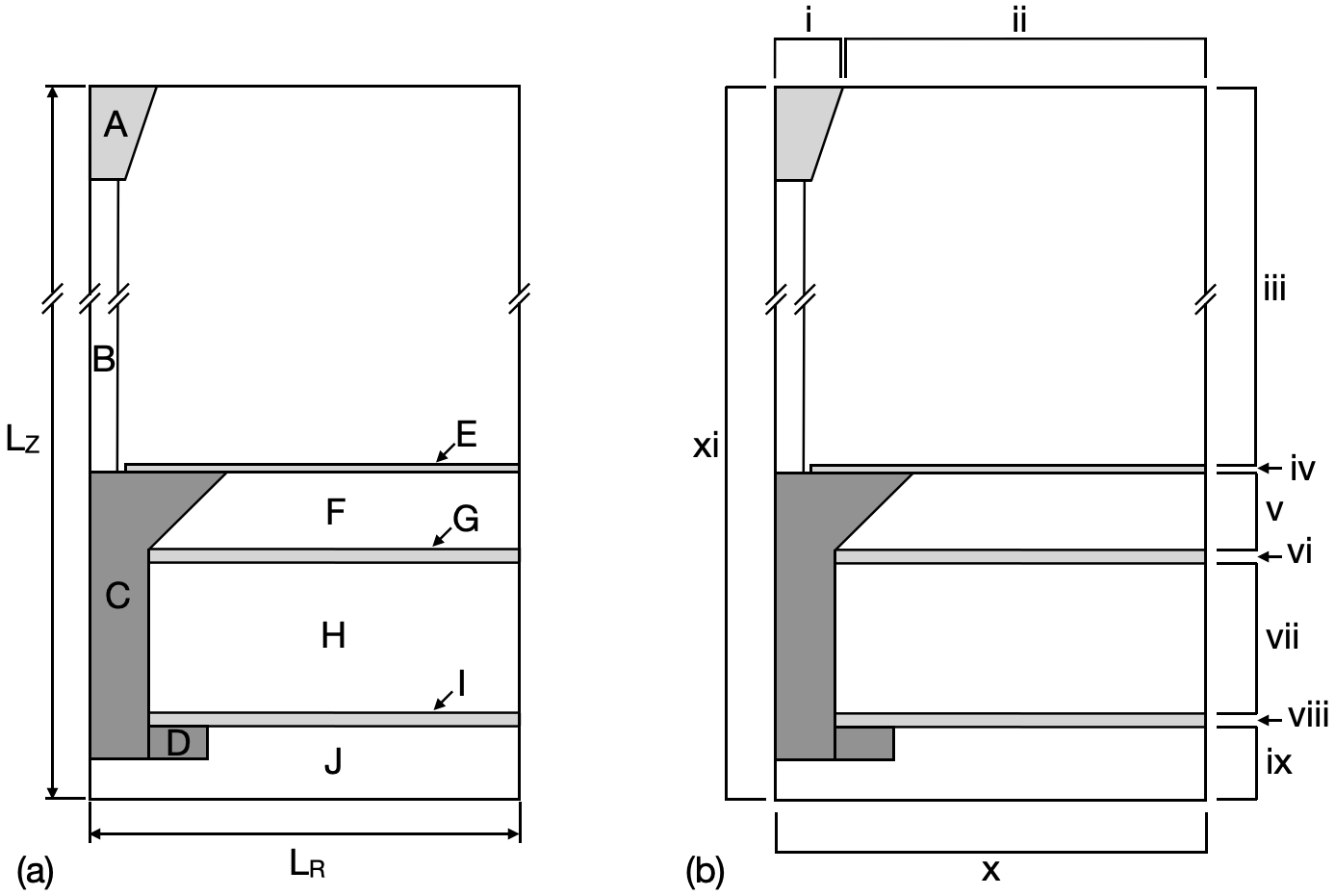}
   \caption{Fastener simulation layout. (a) Physical components, A:
     Electrode, B: Pre-heated arc, C: Fastener shank, D: Fastener
     collar, E: Dielectric layer, F: CFRP {panel}, G: Dielectric
     layer, H: CFRP {panel}, I: Dielectric layer, J: Air. Illustration
     shows a cross-section of the axisymmetric simulation. The
     computational domain extends $\mathrm{L_{R}}$=50.8~mm and
     $\mathrm{L_{Z}}$=62~mm in the radial and axial directions
     respectively. (b) Simulation boundary conditions. Boundary
     conditions (i-xi) given in
     Table~\ref{table:Validation-evans-boundary-table}.}
  \label{fig:validation-evans-component-diagram}
\end{figure}

In this section the multi-physics methodology is applied to an
aerospace fastener configuration used in experimental testing by
Evans~\cite{evans2018thesisCharacterisation}. This can validate the
suitability of the approach outlined in this paper for modelling high
current flow through a complex fastener geometry, comprising a number
of materials with different electrical and thermal properties in
contact. Specifically, a high conductivity titanium fastener
surrounded by successive horizontal layers of carbon fibre and
dielectric is modelled. The methodology used in this paper has
previously been validated against experimental results for lightning
strikes on thin aluminium and carbon composite panels by Millmore and
Nikiforakis~\cite{millmore2019multi}. The experiment of
Evans~\cite{evans2018thesisCharacterisation}, used in this section,
investigates the electrical resistance and current distribution in a
countersunk fastener assembly under attachment of an arc with a 50kA
peak current density, an experimental input which is representative of
a lightning strike. The countersunk head fastener design is typically
used in practice to maintain the smooth profile of the external
aircraft skin. An interference fit of the fastener with the
surrounding CFRP and Glass Fibre Reinforced Polymer (GFRP) layers is
used in the experiment and is replicated in this simulation. An
axisymmetric simulation models the physical experiment in which a
cylindrical fastener is surrounded by a square carbon fibre
panel. Return fasteners are equally spaced along the outer edge of the
panel in the experiment, justifying the axisymmetric approach. A
two-dimensional cross-section of the simulation set-up is given in
figure~\ref{fig:validation-evans-component-diagram}.

Figure~\ref{fig:validation-evans-component-diagram}\,(a) identifies
individual fastener material components using the labels C-I. In
addition, the electrode and pre-heated arc are identified using labels
A and B for reference and the ambient air below the fastener assembly
is labelled J. The radial, $\mathrm{L_{R}}$, and axial,
$\mathrm{L_{Z}}$, lengths of the computational domain are also shown,
where in this work $\mathrm{L_{R}}$=50.8~mm and
$\mathrm{L_{Z}}$=62~mm.  The countersunk titanium fastener, labelled C
in figure~\ref{fig:validation-evans-component-diagram}\,(a), has a
shank diameter of 6.35~mm. The titanium retaining nut, labelled D, has
an outer diameter of 12.4~mm. In the experiment of Evans, a high
voltage electrode is placed in contact with the fastener head. To
replicate this in the present work, in which an electrode is placed
40~mm above the fastener head, a thin dielectric layer (labelled E) is
positioned between the arc and the CFRP panel. This dielectric layer,
which is 0.6~mm thick, ensures that the radially expanding arc
maintains contact only with the fastener head throughout the
simulation. The upper panel, labelled F, is 2.032~mm thick and is
directly in contact with the fastener head. To electrically isolate
the upper panel from the fastener shank and from the lower panel, a
further 0.844~mm thick GFRP dielectric layer, labelled G, is placed
between the two panels. The lower carbon fibre panel, labelled H in
figure~\ref{fig:validation-evans-component-diagram}\,(a) is in
electrical contact with the fastener shank only and is 6.096~mm
thick. To electrically isolate the fastener collar from the lower
panel, a further dielectric layer is used, labelled I, and is 0.5~mm
thick. Above the fastener head, a 4~mm wide plasma arc is initially
defined with a temperature of 8000~K, this is labelled B in
figure~\ref{fig:validation-evans-component-diagram}\,(a). This
pre-heated region is necessary since the breakdown of the air, forming
the initial plasma arc, is not modelled in this framework. This region
is given a sufficient temperature such that a conductive path is
formed between the electrode and the substrate, based on the approach
of Chemartin et al.~\cite{chemartin2011modelling}. Larsson et
al.~\cite{larsson2000lightning} test a number of initial high
temperature (pre-heated) columns up to 20,000K and conclude that the
values within the preheated region do not affect the overall evolution
of the plasma arc. Due to the lack of available equations of state for
GFRP, these layers are approximated in this work using a PMMA equation
of state. For simulating the isotropic approximation to CFRP used in
this work, an electrical conductivity of $\sigma = 8872$~Sm$^{-1}$ is
used, as defined by Evans~\cite{evans2018thesisCharacterisation}.  At
the electrode, a modified `Component A' electrical current waveform,
as defined in ARP 5412B~\cite{ARP5412B}, is applied. For this test,
the input current waveform has a peak current of 50\,kA and is given
by,
\begin{equation}
I\left(t\right) = I_{0}\left(e^{-\alpha t} - e ^{-\beta t} \right)\left(1 - e^{-\gamma t}\right)^{2}
\label{eq:CurrentComponentB}
\end{equation}
where, $\alpha$ = 51,000 s$^{-1}$, $\beta$ = 90,000 s$^{-1}$,
$\gamma$ = 5,423,540 s$^{-1}$ {and} $I_{0}$ = 243,000\,A.
Outside of the plasma arc, the unionised gas is initialised using
ambient ideal gas conditions.

Figure~\ref{fig:validation-evans-component-diagram}\,(b) highlights the
boundary conditions applied in the simulation by labelling eleven
regions of interest, i-xi. Conditions are required for the conserved
variables, $\mathbf{q}$, the electrical potential, $\phi$, as well as the
radial and axial components of magnetic field, $A_{r}$ and $A_{z}$, as defined in
table~\ref{table:Validation-evans-boundary-table}.

\begin{table}
\begin{center}
\begin{tabular}{|c||c|c|c|c|}
\hline
Boundary & \textbf{q} & $\phi$ & $A_r$ & $A_Z$ \\
\hline
\hline
i & Transmissive &$\frac{\partial \phi}{\partial \eta}=-\frac{1}{\sigma}\frac{I(t)}{\pi r^{2}_{c}}$ & $A_{r}=0$ & $\frac{\partial A_{z}}{\partial \eta} = 0$ \\
\hline
ii & Transmissive  & $\frac{\partial \phi}{\partial \eta}=0$ & $A_{r}=0$ & $\frac{\partial A_{z}}{\partial \eta} = 0$ \\
\hline
iii & Transmissive & $\frac{\partial \phi}{\partial \eta}=0$ & $\frac{\partial A_{r}}{\partial \eta} = 0$ & $A_{z} = 0$ \\
\hline
iv & Transmissive  & $\frac{\partial \phi}{\partial \eta}=0$ & $\frac{\partial A_{r}}{\partial \eta} = 0$ & $A_{z} = 0$ \\ 
\hline
v & Transmissive & $\phi=0$ & $\frac{\partial A_{r}}{\partial \eta} = 0$ & $A_{z} = 0$ \\
\hline
vi & Transmissive & $\frac{\partial \phi}{\partial \eta}=0$ & $\frac{\partial A_{r}}{\partial \eta} = 0$ & $A_{z} = 0$ \\ 
\hline
vii & Transmissive & $\phi=0$ & $\frac{\partial A_{r}}{\partial \eta} = 0$ & $A_{z} = 0$ \\
\hline
viii & Transmissive & $\frac{\partial \phi}{\partial \eta}=0$ & $\frac{\partial A_{r}}{\partial \eta} = 0$ & $A_{z} = 0$ \\
\hline
ix & Transmissive & $\phi=0$ & $\frac{\partial A_{r}}{\partial \eta} = 0$ & $A_{z} = 0$ \\ 
\hline
x & Transmissive & $\frac{\partial \phi}{\partial \eta}=0$ & $\frac{\partial A_{r}}{\partial \eta} = 0$ & $A_{z} = 0$ \\
\hline
xi & Symmetry & $\frac{\partial \phi}{\partial \eta}=0$ & $A_{r}=0$ & $\frac{\partial A_{z}}{\partial \eta} = 0$ \\
\hline
\end{tabular}
\caption{Simulation boundary conditions for the conserved
  variables, $\textbf{q}$, electric potential,
  $\phi$, the radial and axial components of magnetic potential,
  $A_{r}$ and $A_{z}$. Boundary indices in the first column
  correspond to
  figure~\ref{fig:validation-evans-component-diagram}(b).}\label{table:Validation-evans-boundary-table}
\end{center}
\end{table}

\begin{figure}[!ht]
  \centering
   \includegraphics[width=0.4\textwidth]{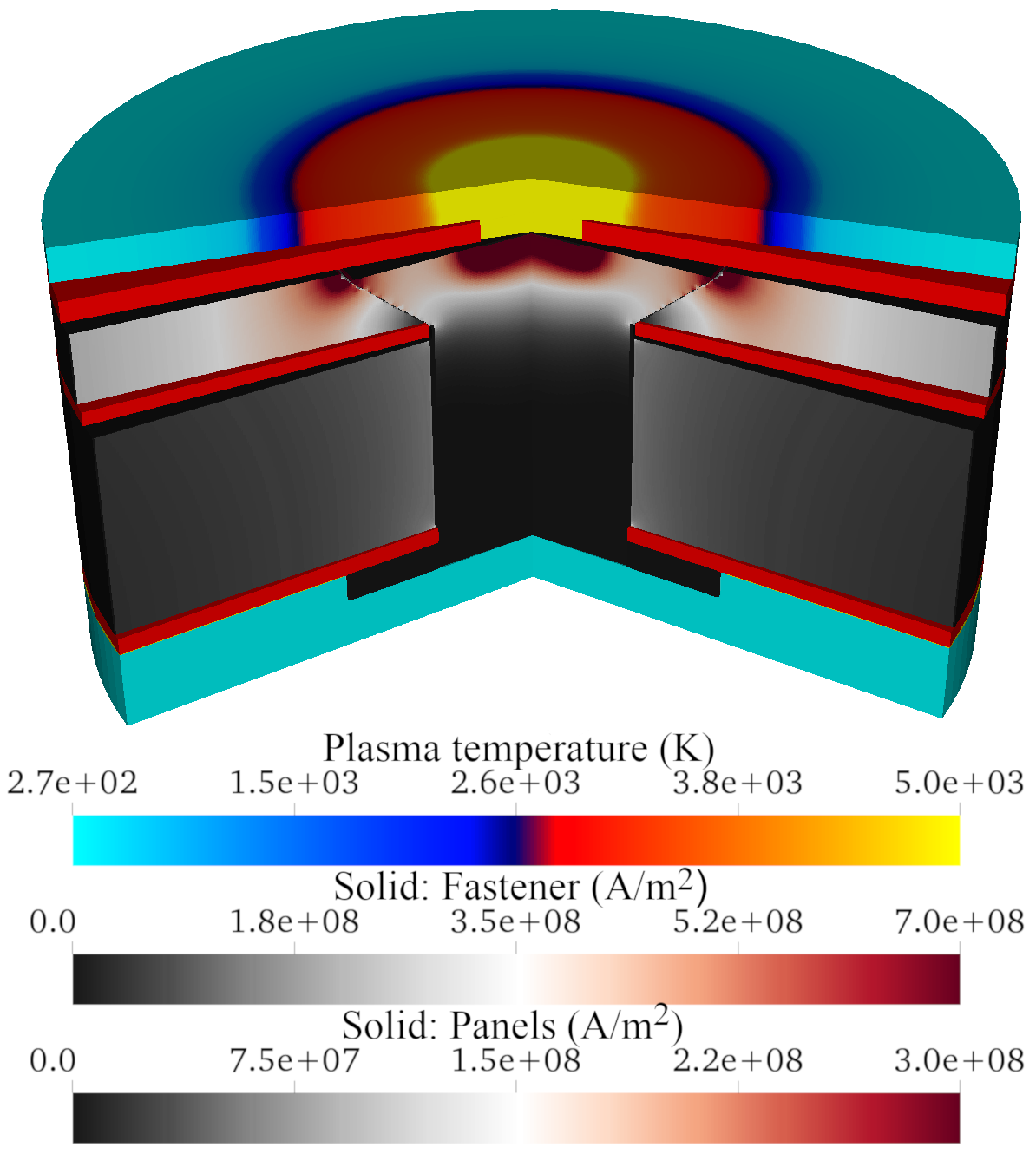}
  \caption{Fastener current density and plasma temperature at a time
    of 30 $\mu$s. The plasma region has been
    truncated in this figure to allow greater detail of the fastener
    to be shown.}
  \label{fig:validation-evans-snapshot25}
\end{figure}

Figure~\ref{fig:validation-evans-snapshot25} shows a snapshot of this
test case at a time of 30~$\mu$s. This plot shows current density
within the composite substrate materials and the temperature within
the plasma arc; the dielectric layers are shown in red. The regions of
highest current density are highlighted as being at the attachment
point at the top of the countersunk fastener bolt head, at the outer
radial tip of the countersunk head and along the upper panel.  It is
clear that the preferred path for current flow is through the upper
panel, close to the surface.  A localised peak in current density
magnitude is visible at the upper radial edge of the countersunk
fastener head, at the interface with the upper carbon composite
panel. The difference in electrical conductivity between the titanium
fastener and the low conductivity panels results in a preference for
the current path to remain in the higher conductivity fastener for as
long as possible. The shape of the countersunk fastener head outer
edge acts to promote a gradient in current density magnitude between
the top and bottom of the upper carbon composite panel, at the
interface with the fastener. Together with the small axial cross
sectional area of the fastener at the tip of the countersunk head,
this results in a local peak in current density magnitude, and could
cause issues with high localised temperature and pressure through
Joule heating. The experimental post-test specimen in Kirchdoerfer
et al.~\cite{kirchdoerfer2017} shows the greatest arc-induced erosion
of the CFRP panel at this location in their countersunk fastener case
using a 40~kA peak current waveform. At the bottom of the fastener in
figure~\ref{fig:validation-evans-snapshot25}, the electrical isolation
of the fastener nut by the bottom dielectric layer results in
extremely low current density magnitude values in the fastener shank
and nut at this location.

\begin{figure}[!ht]
  \centering
   \includegraphics[width=0.45\textwidth]{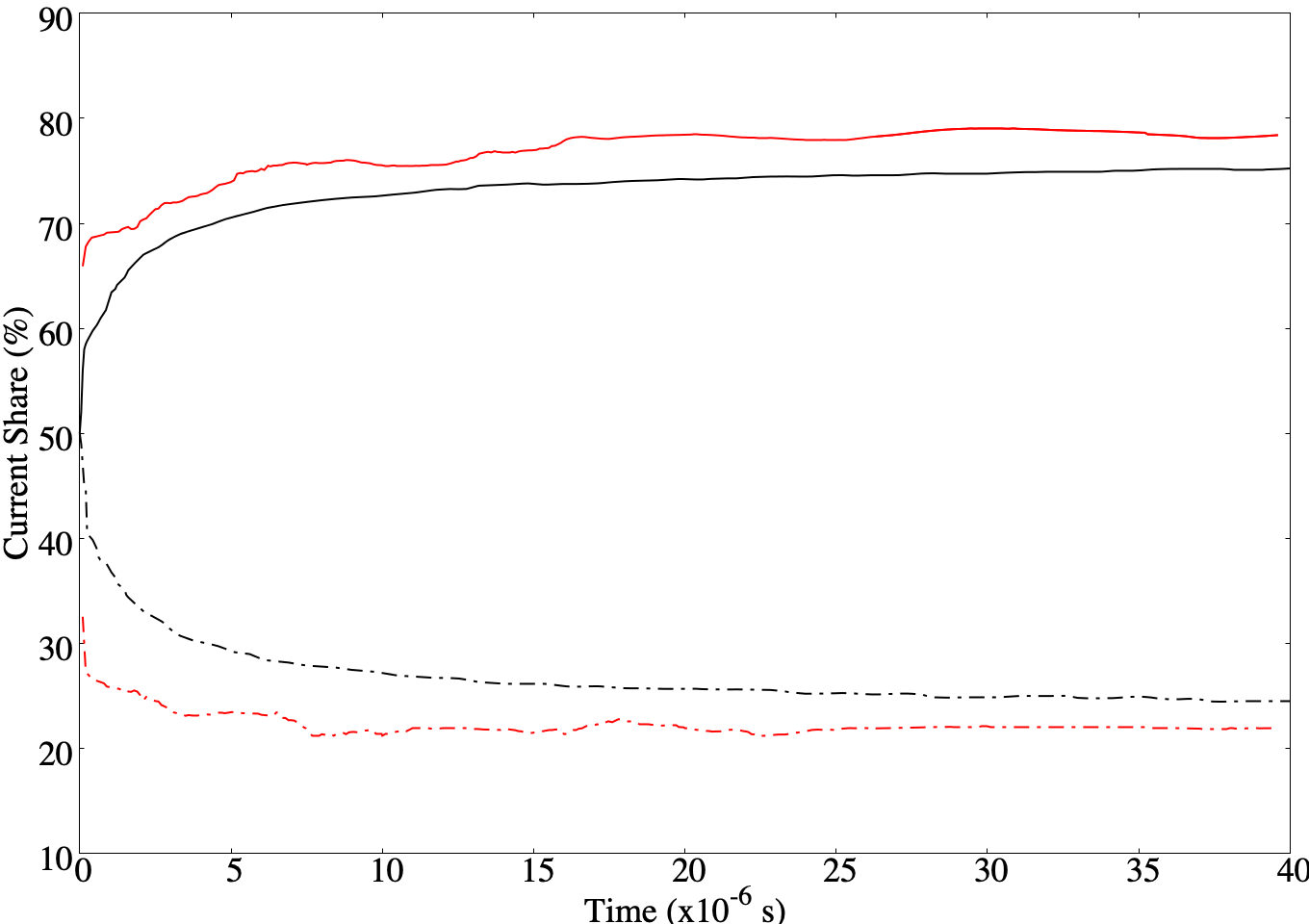}
  \caption{Percentage of current passing through the upper CFRP
    panel via the countersunk head (solid lines)
    and through the lower CFRP panel via the fastener
    shank (dash-dot line). Numerical results are shown
      with red lines, and the experimental results of
    Evans~\cite{evans2018thesisCharacterisation} with black lines.}
  \label{fig:validation-evans-current-share}
\end{figure}

In the experiment of Evans~\cite{evans2018thesisCharacterisation}, the
current flow through the upper and lower composite panels are measured
independently using alternating cut-out sections at the outer edge of
each panel so that each return fastener contacts only the upper or
lower panel. The relative proportion of total current passing through
the upper and lower panels is then measured over time. The electrical
current share is computed from the simulation results by integrating
current density along a line normal to the current flow
direction. This line is taken at a radius of 12~mm; at which point the
current density streamlines were effectively purely radial.  The
measurements of Evans show that for an interference fit fastener, over
three quarters of the total current passes through the upper panel,
identifying this route as the preferred current
path. Figure~\ref{fig:validation-evans-current-share} compares the
experimental and simulation results for the relative share of total
current passing through the upper and lower composite panels over the
first 40~$\mu$s.

Comparing the simulation and the experiment in
figure~\ref{fig:validation-evans-current-share}, there is a similar
distribution of current between the upper and lower panels in both
cases. The significantly higher proportion of current flowing through
the upper panel reflects the large area of contact between the upper
panel and the fastener head, as well as the preference for current to
travel via the least resistive route to ground. The percentage of
current passing through the upper panel is slightly greater in the
simulation than in the measurements of
Evans~\cite{evans2018thesisCharacterisation}. This may in part be due
to the difference in current input at the top of the fastener, with
the simulation using an attached arc, rather than a point source. The
radial expansion of the arc increases the contact area between the
simulated arc and fastener head. Once the arc radius has increased
past the radius of the fastener shank, a significant proportion of the
current density input is close to the outer radius of the fastener
head, resulting in a greater current flow across the short distance
between the top of the fastener head an the upper panel.  However, the
close match in behaviour over time does confirm that this
computational approach is capable of simulating the transient current
flow in a complex fastener geometry comprising a number of distinct
layers of materials with differing electrical and thermal properties.


\section{Fastener design sensitivity studies}
\label{sec:SensitivityStudies}

This section considers the influence of fastener design choices in the
distribution of current flow and associated pressure and temperature
increases in the fastener assembly. In
section~\ref{sec:FastenerGeometry} the location of dielectric layers
in the fastener assembly is considered. Dielectric layers can be
included in fastener assemblies through judicious design to control
current flow away from sensitive components, or as a result of
component sealant and resin use. The inclusion of a clearance gap
between the fastener shank is also considered in this section by
comparison with an interference fit type
fastener. Section~\ref{sec:FastenerPreHeating} develops the clearance
fit gap analysis by considering the necessity in the present numerical
modelling method for pre-heating of the clearance fit gap to promote
an electrically conductive path across the gap. The effect of
clearance gap width is then considered in
section~\ref{sec:FastenerShankWidth}, both with and without
pre-heating.  All simulations in this section use the current
described by equation~(\ref{eq:CurrentComponentB}) with parameters as
defined in ARP 5412B~\cite{ARP5412B}, $\alpha$ = 11,354 s$^{-1}$,
$\beta$ = 647,265 s$^{-1}$, $\gamma$ = 5,423,540 s$^{-1}$ {and}
$I_{0}$ = 43,762\,A.

\subsection{The effect of dielectric layers and fastener clearance fitting}
\label{sec:FastenerGeometry}

The multi-physics methodology outlined in
section~\ref{sec:MathematicalModel} allows not only the mechanical,
thermodynamic and electrodynamic evolution of an aerospace fastener
subject to lightning attachment to be captured, but also for the
sensitivity of these properties to changes in fastener design to be
assessed.

\begin{figure}[!ht]
  \centering
   \includegraphics[width=0.45\textwidth]{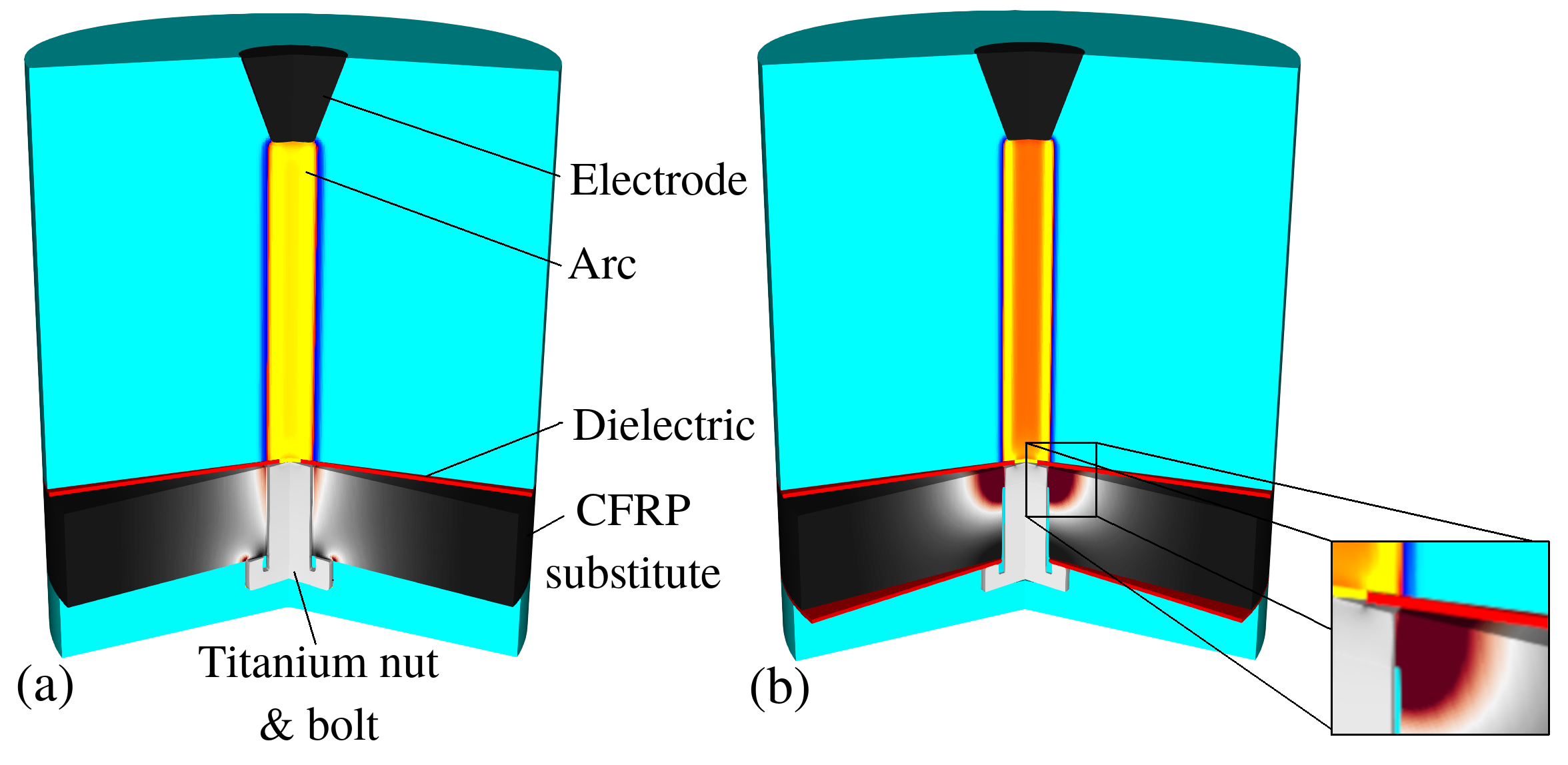}
   \caption{Annotated schematic of two different fastener
     geometries. (a) A dielectric layer above the panel prevents
     direct contact between the panel and the arc and an interference
     fit exists between the fastener and the panel. (b) Dielectric
     layers are placed on both the top and bottom of the panel,
     preventing direct contact between the fastener nut and the
     underside of the panel, and a clearance fit exists between the
     fastener and the panel. The insert shows an enlarged section
     highlighting the clearance gap.}
  \label{fig:Case7and8InitialAnnotated}
\end{figure}

Figure~\ref{fig:Case7and8InitialAnnotated} shows an example of two
different idealised aerospace fastener configurations. Both fasteners
comprise a titanium nut and bolt and a single carbon composite
panel. An electrode is placed 40~mm above the substrate surface, and a
pre-heated region exists between this and the substrate.

The first fastener design, shown in
figure~\ref{fig:Case7and8InitialAnnotated}\,(a), has a dielectric
coating layer on the upper face of the carbon composite panel. This
layer, which is 0.5~mm thick, represents, for example, a painted
surface or other coating. The titanium bolt is 6~mm in diameter and is
in direct contact with a composite panel of thickness 12~mm,
representing an interference-type fit. A titanium nut, of diameter
12~mm and thickness 2~mm, is in direct contact with the lower face
of the carbon composite panel. The second idealised fastener design,
figure~\ref{fig:Case7and8InitialAnnotated}\,(b) has dielectric layers at
both upper and lower surfaces of the carbon composite panel, each
of thickness 0.5~mm. A clearance fit air gap between the titanium bolt
and the surrounding carbon composite panel of width 0.3~mm is defined
in the second fastener design. This is shown as a enlarged region in
figure~\ref{fig:Case7and8InitialAnnotated}\,(b). The only area of direct
contact between the fastener bolt and the carbon composite panel is a
2.5~mm length directly below the top dielectric layer.

In both fastener designs, a pin hole punctures the radial centre of
the upper dielectric layer. This pin hole is a technique frequently
used in lightning experiments, and is used in this case to initialise a
conductive path between the plasma arc and the titanium fastener. In
practice, this pin hole may crudely represent a small region of
dielectric ablation during the initial attachment of the lightning
strike. The diameter of the hole at the centre of the upper dielectric
layer is reduced from 6~mm to 3~mm in the latter configuration. The
final difference between the two configurations is the introduction of
a narrow strip of non-ionised air between the fastener shank and the
panel. In order to reduce the computational resolution overhead in
these idealised fastener designs, the thread between the bolt and nut
is neglected.

\begin{figure}[!ht]
  \centering
   \includegraphics[width=0.45\textwidth]{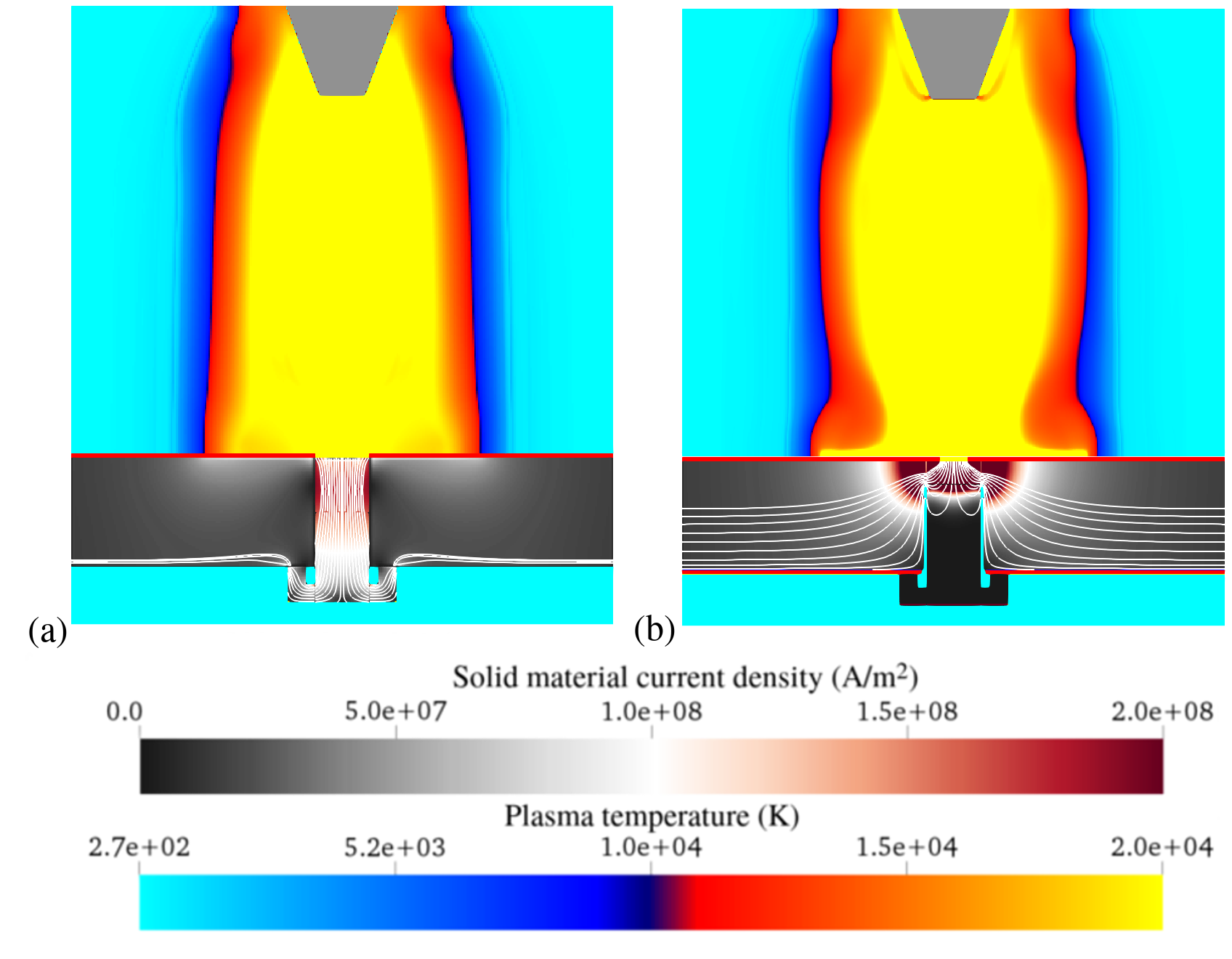}
   \caption{ Showing the temperature profile within the plasma arc,
     and current density magnitude within the substrate, after
     20~$\mu$s for the two fastener designs shown in
     figure~\ref{fig:Case7and8InitialAnnotated}.  Current density
     streamlines are also shown in the substrate, as white lines.  In
     the case of direct contact between the titanium nut and the
     substrate in (a), despite the interference fit between between
     the fastener and the substrate, the preferred path for current is
     through the nut.  When a second dielectric layer is introduced,
     in (b), the current now flows directly from the fastener to the
     substrate.  Although the clearance fit results in the the
     majority of the current entering the substrate where it is in
     contact with the fastener, it subsequently spreads out as it
     travels radially outwards.}
  \label{fig:Case7and8Streamlines}
\end{figure}

Figure~\ref{fig:Case7and8Streamlines} shows the current density
profiles in the substrate, and temperature in the plasma arc, for the
two fastener configurations shown in
figure~\ref{fig:Case7and8InitialAnnotated} at a time of
20~$\mu$s. Current density streamlines are also shown in the
substrate materials. As expected, for the first fastener configuration
in figure~\ref{fig:Case7and8Streamlines}\,(a), the main electrical
current path is along the titanium fastener bolt, through the titanium
nut and into the lower face of the carbon composite panel to the
ground location at the outer radius of the panel. The second fastener
configuration, figure~\ref{fig:Case7and8Streamlines}\,(b), is
electrically insulated between the titanium nut and the panel, and due
to the clearance fit used in this case, the preferred path for current
is through the small region of contact between the fastener and the
panel.

This results in high current density close to the top of the fastener,
seen in figure~\ref{fig:Case7and8Streamlines}\,(b), and subsequent
current flow into the substrate at this point.  It is clear from the
current density streamlines that the subsequent flow through the panel
is then spread over a wider area than in the case where the lower dielectric layer
is not used.

The reduction in the hole diameter of the upper dielectric and changes
in current density magnitude also influence the shape of the arc above
the fastener. A local radial restriction, or `pinching', of the plasma
arc is evident in figure~\ref{fig:Case7and8Streamlines}\,(b),
immediately above the attachment point; this pinching is not evident
to the same extent in figure~\ref{fig:Case7and8Streamlines}\,(a). The
ability for the present numerical method to identify the dependence of
arc characteristics on the configuration, material choice and layering
of the substrate highlights the advantage of a fully coupled
system. This coupling was demonstrated and validated against
experimental results in Millmore and
Nikiforakis~\cite{millmore2019multi}.

\begin{figure}[!ht]
  \centering
   \includegraphics[width=0.5\textwidth]{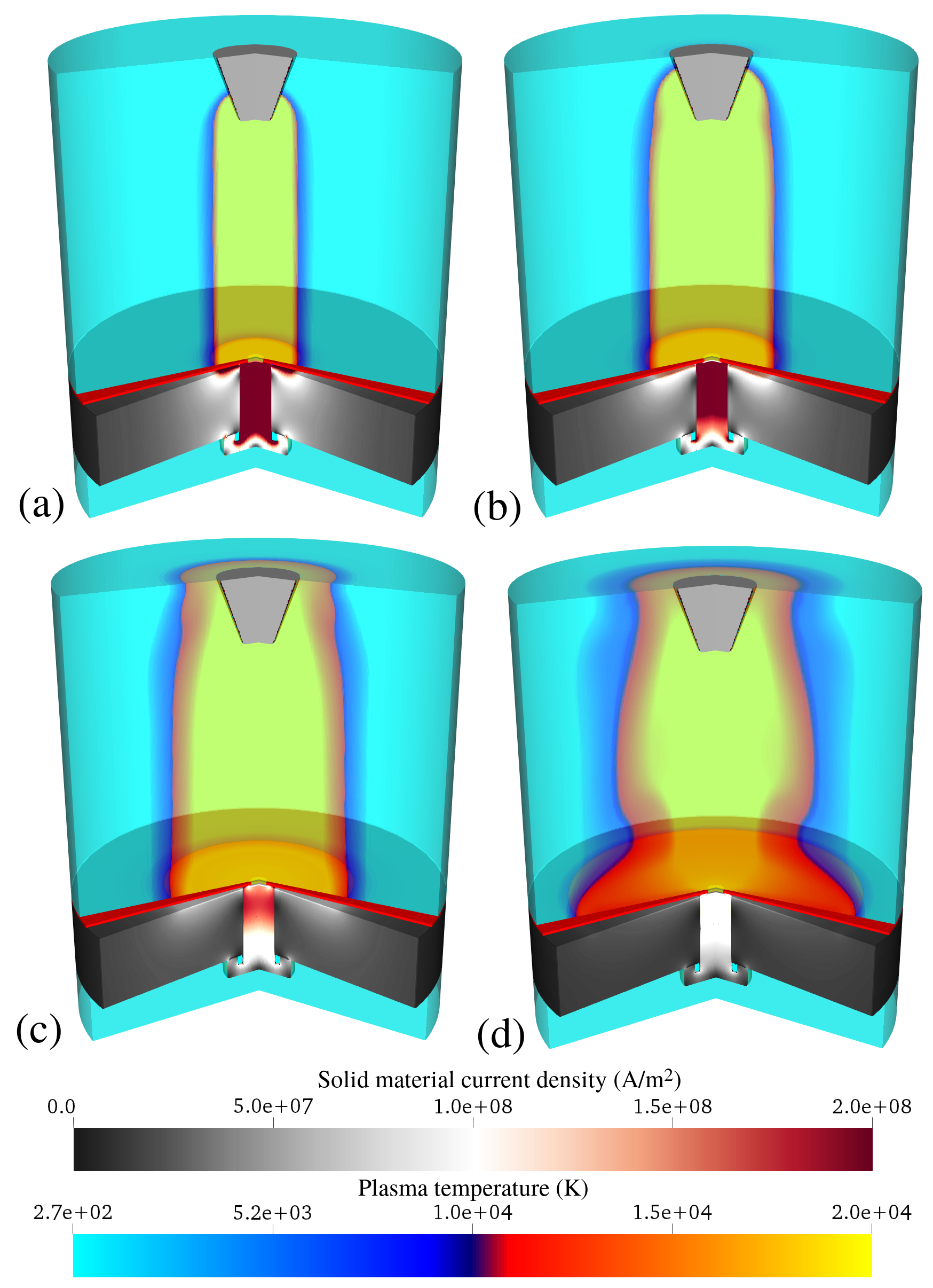}
   \caption{ Evolution of current density magnitude in the substrate
     and temperature in the plasma arc over the first 40~$\mu$s
     following plasma arc attachment for the fastener configuration
     illustrated in
     figure~\ref{fig:Case7and8InitialAnnotated}\,(a). Times shown are
     (a) 5~$\mu$s, (b) 10~$\mu$s, (c) 20~$\mu$s and (d) 40~$\mu$s.
     The radial expansion of the arc is clearly visible, with some
     pinching effects at later times.  The reduction in current
     density magnitude over time within the fastener follows the input
     current profile at the electrode, and the higher values
     throughout the fastener, compared to the panel, show this being
     the preferred path for current flow.}
  \label{fig:Case7_Snapshots}
\end{figure}

The thermal, mechanical and electrodynamic development of the first
fastener configuration over the first 40~$\mu$s is shown in
figure~\ref{fig:Case7_Snapshots}, with snapshots shown at 5~$\mu$s,
10~$\mu$s, 20~$\mu$s and 40~$\mu$s. The increase in the radial extent
of the plasma arc is evident, as is the rise and fall in current
density in the fastener over time as the current input in the
electrode rises and falls according to
equation~\ref{eq:CurrentComponentB}. Figure~\ref{fig:Case7_Snapshots}\,(c)
corresponds with the current density streamlines shown in
figure~\ref{fig:Case7and8Streamlines}\,(a) .

\begin{figure}[!ht]
  \centering
   \includegraphics[width=0.5\textwidth]{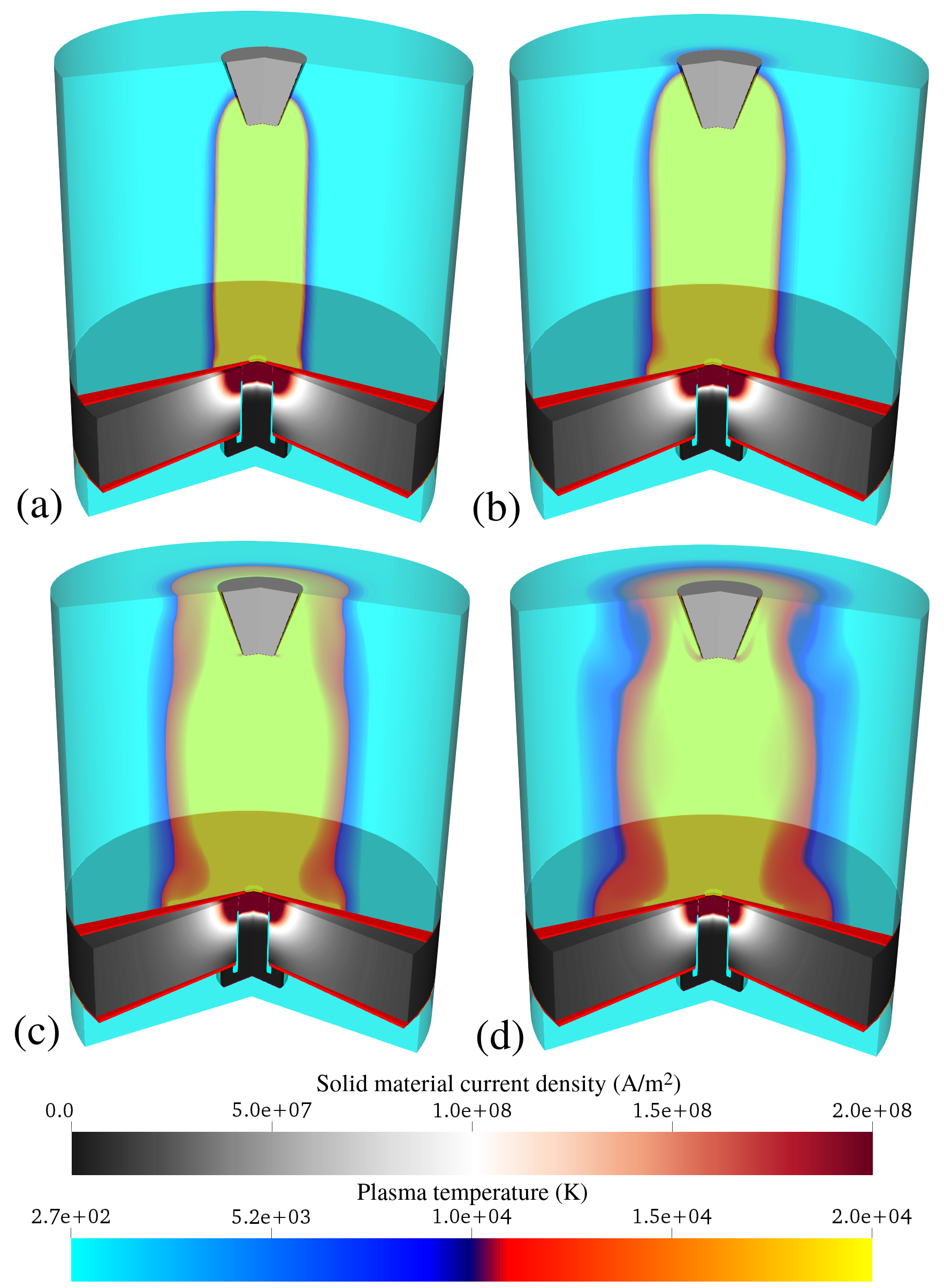}
   \caption{ Evolution of current density magnitude in the substrate
     and temperature in the plasma arc over the first 40~$\mu$s
     following plasma arc attachment for the fastener configuration
     illustrated in
     figure~\ref{fig:Case7and8InitialAnnotated}\,(b). Times shown are
     (a) 5~$\mu$s, (b) 10~$\mu$s, (c) 20~$\mu$s and (d) 40~$\mu$s. The
     pinching of the plasma arc is much more visible in this case,
     than in figure~\ref{fig:Case7_Snapshots}, and is apparent from 10
     $\mu$s onwards.  Due to the small contact region between the
     fastener and the panel, the current density magnitude remains
     high at the attachment point.}
  \label{fig:Case8_Snapshots}
\end{figure}

The thermal, mechanical and electrodynamic development of the second
fastener configuration over the first 40~$\mu$s is shown in
figure~\ref{fig:Case8_Snapshots}, at times directly comparable to
figure~\ref{fig:Case7_Snapshots}. In this case,
figure~\ref{fig:Case8_Snapshots}\,(c) corresponds with the current
density streamline plot shown in
figure~\ref{fig:Case7and8Streamlines}\,(b).

The increase in arc radius over time is again evident, as is the
difference in arc shape between the two fastener configurations. The
current density at the region of direct contact between the fastener
bolt and the carbon composite panel remains high throughout the
evolution, and this then leads to higher local pressures at the
interface between the fastener and panel through increased Joule
heating. The interface between the high conductivity titanium fastener
and the lower conductivity panel is one area of particular concern in
fastener design to reduce the possibility of local breakdown in
material integrity. Such behaviour is more likely with the higher
current density magnitude over a limited spatial distribution around
the upper region of the fastener bolt in the case shown in
figure~\ref{fig:Case8_Snapshots}, in comparison to
figure~\ref{fig:Case7_Snapshots}. As this is associated with
higher levels of Joule heating and increased pressure rise at the
top of the bolt, figures~\ref{fig:Case7_Snapshots} and~\ref{fig:Case8_Snapshots} highlight the advantage of fastener
designs that maximise the physical contact area of conductive materials across the fastener-panel interface. This approach is evident in the metallic
sleeve designs of Mulazimoglu \& Haylock~\cite{mulazimoglu2011recent}.

The analysis in this section is extended in the next section to assess how changes in
fastener design can effect the temperature and pressure rise in a gap
between the fastener and carbon composite panels. Excessive pressure
rise in internal fastener gaps is a potential cause of unwanted
outgassing events and electrical sparking in
fasteners~\cite{kirchdoerfer2017,kirchdoerfer2018cth,teulet2017energy}.


\subsection{The effect of clearance gap pre-heating}
\label{sec:FastenerPreHeating}

In practice, a fastener assembly may contain internal gaps between,
for example, the bolt and the composite panels, or between the bolt
and nut. Under the high current input conditions of a
lightning strike, ionisation of the gas within the internal voids
may occur. Ionisation of the trapped gas can establish an electrically
conductive path across the void, which can lead to changes in the
current distribution of the assembly and significant increases in the gap 
pressure. In this section, a method for modelling the
development of a conductive path across internal voids is
considered and a pre-heating approach to initialise this is investigated.

Building on the initial application of the multi-physics methodology
to enable assessment of changing thermal and mechanical behaviour
from variations in fastener design and material choice. We now extend this analysis to the internal gap 
between the fastener bolt, nut and adjacent panel. 
To enable this extension, a modification is made
to the fastener design in the previous section to include a bolt
head. It is recognised that in practical fastener designs, a
counter-sunk fastener head is often used to enable a flush panel
surface and to maximise direct contact with the surrounding
substrate. The idealised fastener chosen as an exercise in this
section has, however, been designed to limit the direct contact
between the fastener bolt and the panel to a region around the bolt
head. This is expected to result in a localised region of high current
density, with an associated rise in the pressure and temperature
through Joule heating. This would therefore make a poor aerospace
fastener design in reality but satisfactorily serves here as an
idealised model to investigate numerically establishing a conductive
path across internal voids. The titanium bolt has a clearance fit with
the surrounding panel. The gap is initially assumed
to contain air at ambient conditions. In later testing in this
section, this assumption is altered with the assumption that
electrical breakdown has occurred. 

\begin{figure}[!ht]
  \centering
   \includegraphics[width=0.5\textwidth]{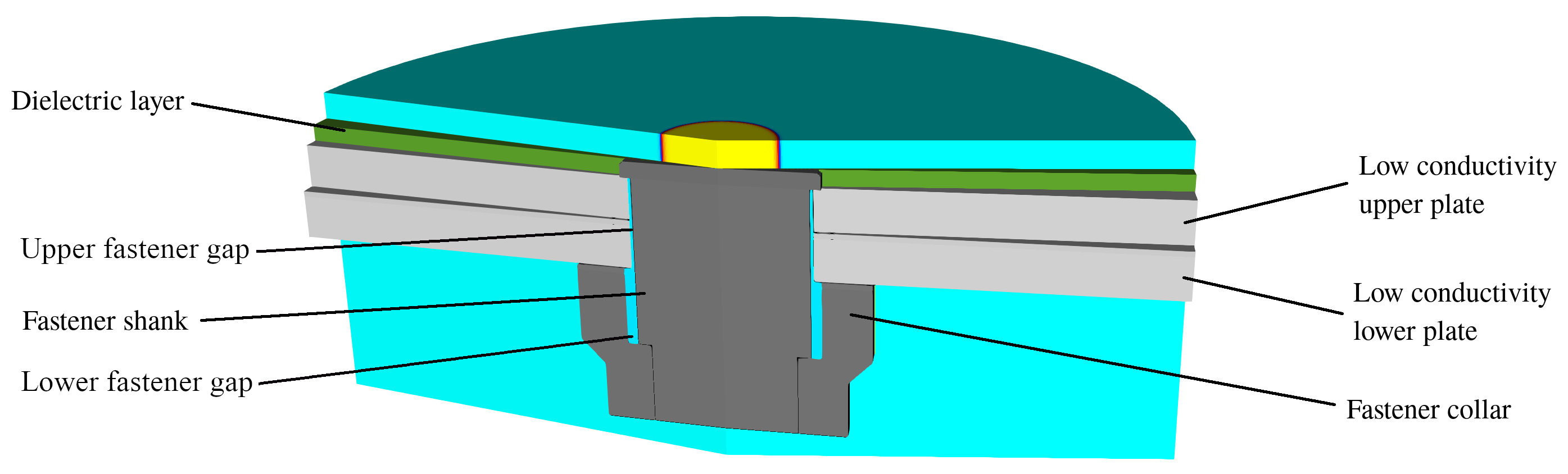}
  \caption{ Schematic of an idealised fastener geometry
      for investigating the behaviour of air gaps between the fastener
      and substrate panels.  Two carbon composite panels are used, and
      the fastener comprises a bolt head, which is in direct contact
      with the upper panel, and a shank with a clearance fit between
      it and the panels.  The fastener collar, however, is in contact
      with the lower panel, and a dielectric coating is considered, as
      in previous tests, on top of the upper panel.}
  \label{fig:BoeingFastener_InitialAnnotated}
\end{figure}

The idealised fastener design used in this section is shown in
figure~\ref{fig:BoeingFastener_InitialAnnotated}. The fastener head
is located above two low conductivity panels, sitting flush
with the upper surface of a dielectric coating. The titanium fastener
has a shank diameter of 6~mm, and this screws into a securing titanium
nut (collar) with an outer diameter of 12~mm. Again, in this
simplified example the fastener thread detail is not considered. The
titanium collar rests on the underside of the lower carbon composite
panel, establishing a direct electrical contact between the collar and
the lower panel. There is a narrow gap between the fastener and the
collar, highlighted as a blue region in
figure~\ref{fig:BoeingFastener_InitialAnnotated}. The width of this
gap (0.38~mm) is larger than the gap that exists between the fastener
bolt and the carbon composite panel (0.1~mm). The carbon composite
substrate is split into two narrower panels, each of thickness
1.7~mm.

The evolution of this system is governed not only by the
electrical conductivity of the various components, but also through
contact resistance between them.  A contact resistance of 1~m$\Omega$
is defined between the two carbon composite panels and a further
contact resistance region of 1~m$\Omega$ is defined between the lower
panel and the titanium collar. The contact resistance is taken as a
typical value from the literature, as used by, for example Chemartin
et al.~\cite{chemartin2013modeling}. In this work it is assumed to be
constant. Mastrolembo~\cite{mastrolembo2017understanding}, however, 
discusses the change in contact resistance that can occur with variations in mechanical loading. 
In the present fastener configuration, this mechanical loading would relate to the tightening torque of the fastener.

The transient current waveform given by
equation~\ref{eq:CurrentComponentB} is defined along the upper domain
boundary and a narrow pre-heated region of the domain at the radial
centre is imposed. This pre-heated region is of sufficient temperature
(8000~K) for ionisation of the plasma. In this first simulation, the
upper and lower fastener gaps are under atmospheric conditions, i.e.\
there is no initial ionised material present. This is intended to
contrast with later simulations in which a pre-heated region is
present close to the top of the gap at the start of the computation.

\begin{figure}[!ht]
  \centering
   \includegraphics[width=0.5\textwidth]{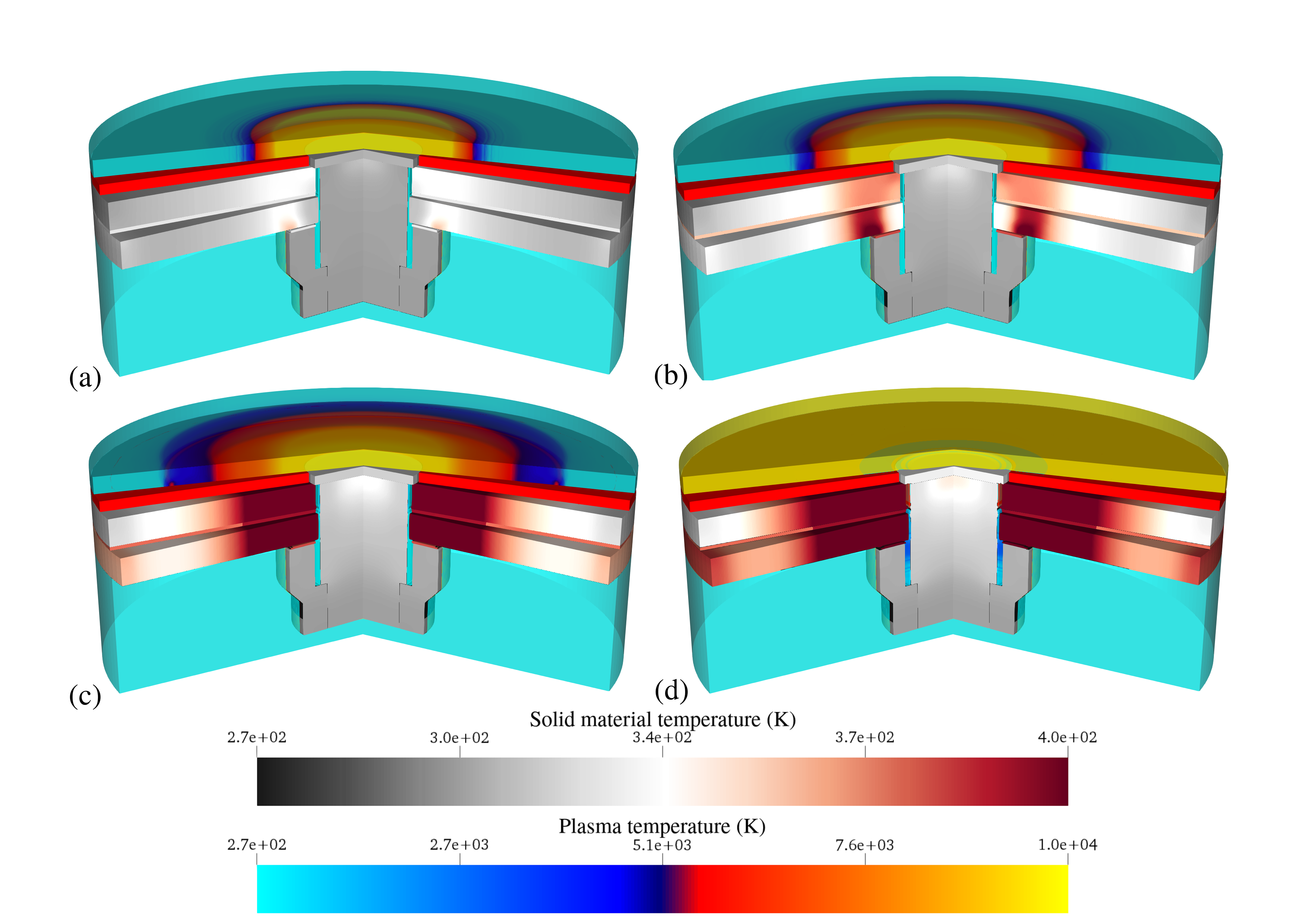}
   \caption{ Temperature evolution over the first 40~$\mu$s after arc
     attachment for the configuration shown in
     figure~\ref{fig:BoeingFastener_InitialAnnotated}. Times shown
     are: (a) 5~$\mu$s, (b) 10~$\mu$s, (c) 20~$\mu$s and (d)
     40~$\mu$s.  It is clear that there is current flow through both
     carbon composite panels, both show a strong temperature rise,
     with the lower panel being at slightly higher temperature than
     the upper panel.}
  \label{fig:BoeingFastener_Snapshots}
\end{figure}

The temperature profiles for the plasma and substrate over the 
first 40~$\mu$s in this configuration is shown in
figure~\ref{fig:BoeingFastener_Snapshots}. A high temperature is
evident in the low conductivity panels, in contrast to the highly
conductive titanium fastener bolt and collar which show little sign of
heating. A rise in temperature occurs in both upper and lower panels,
highlighting two main current paths; through the fastener bolt head
and through direct contact between the collar and lower panel.

\begin{figure}[!ht]
  \centering
   \includegraphics[width=0.45\textwidth]{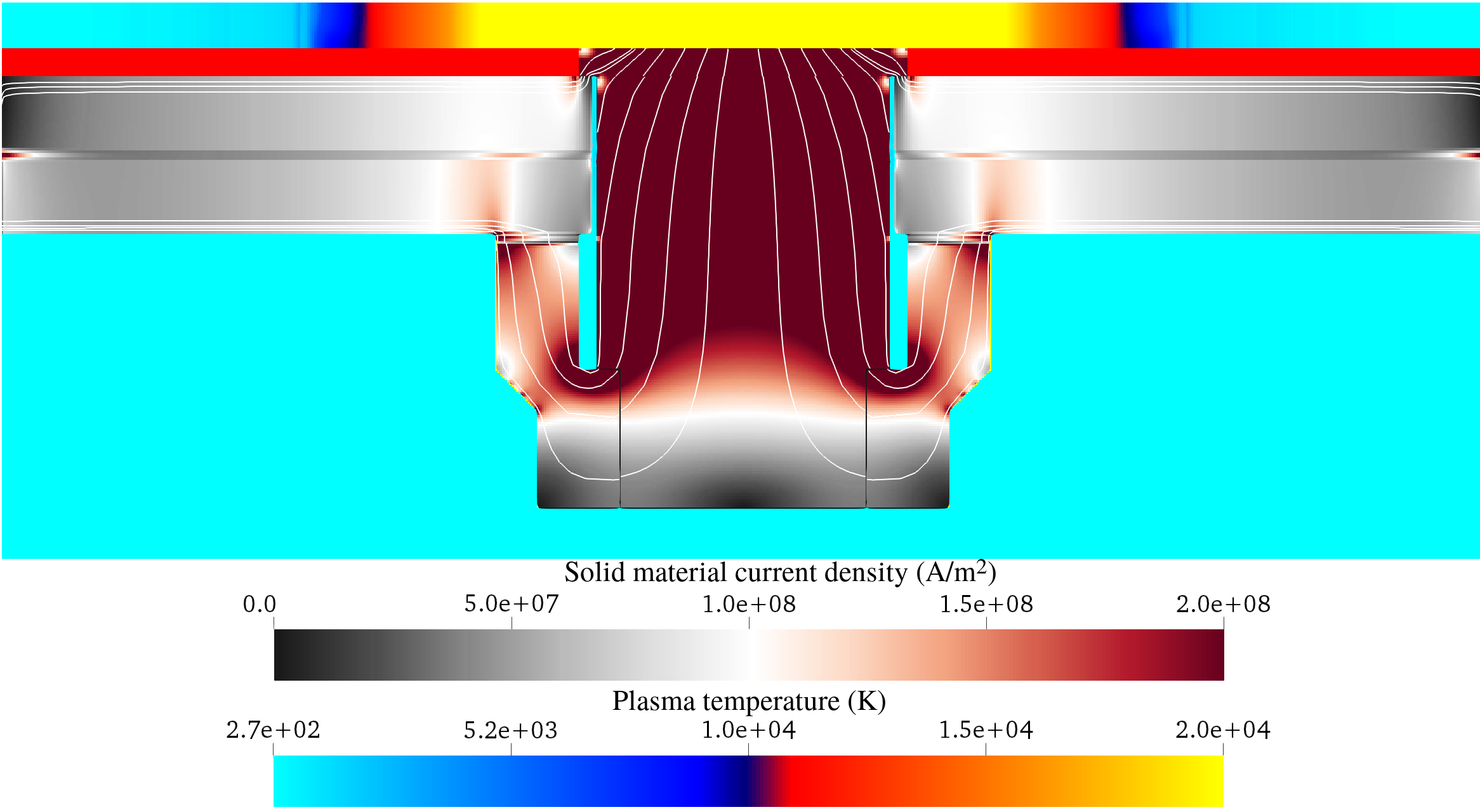}
   \caption{ Current density magnitude and streamlines (white lines)
     within the substrate materials for the fastener configuration
     shown in figure~\ref{fig:BoeingFastener_InitialAnnotated} after
     10~$\mu$s.  It is clear that current passes through both panels
     of carbon composite, though due to the grounding of the lower
     panel, the current density magnitude is greater here.  The rise
     in current density at the outer edges of the panels is due to
     current flow from the upper panel to the lower panel, and hence
     to the ground site.}
  \label{fig:BoeingFastener_Streamlines}
\end{figure}

Plotting the current density magnitude and streamlines, as shown in
figure~\ref{fig:BoeingFastener_Streamlines}, highlights these two
primary current pathways through the fastener geometry. In this
preliminary study, only the lower panel is grounded, the current in
the upper panel passes to the lower panel at the radial extent of the
lower panel. This is visible as a small region of higher current
density magnitude at the outer radial edge of the inter-panel region.

The local increase in current density at the fastener bolt results in
a local pressure and temperature rise from Joule heating. Through
the multi-material boundary conditions at the interface between the
fastener gap and the surrounding materials, this leads to a rise in
these properties within the gap itself.  This is particularly apparent
in the upper-most region of the fastener gap, where conditions are
sufficient for ionisation of the confined air to occur.  Once
this happens, the fastener gap becomes a viable path for current
passage between the fastener and the panel.

Although there is a significant increase in temperature of over
5000~K within the fastener gap over the first 40~$\mu$s, the
corresponding rise in plasma pressure in the cavity is markedly below
published experimental measurements for comparable
  configurations. For example, pressures of 25\,-\,30~MPa (typical)
and 70~MPa (peak) are reported in Kirchdoerfer et
al.~\cite{kirchdoerfer2017} for a comparable peak input current, or
24~MPa\,-\,33~MPa for the 10~mm$^{3}$ volume tested sample in Teulet et
al.~\cite{teulet2017energy}.

\begin{figure}[!ht]
  \centering
   \includegraphics[width=0.3\textwidth]{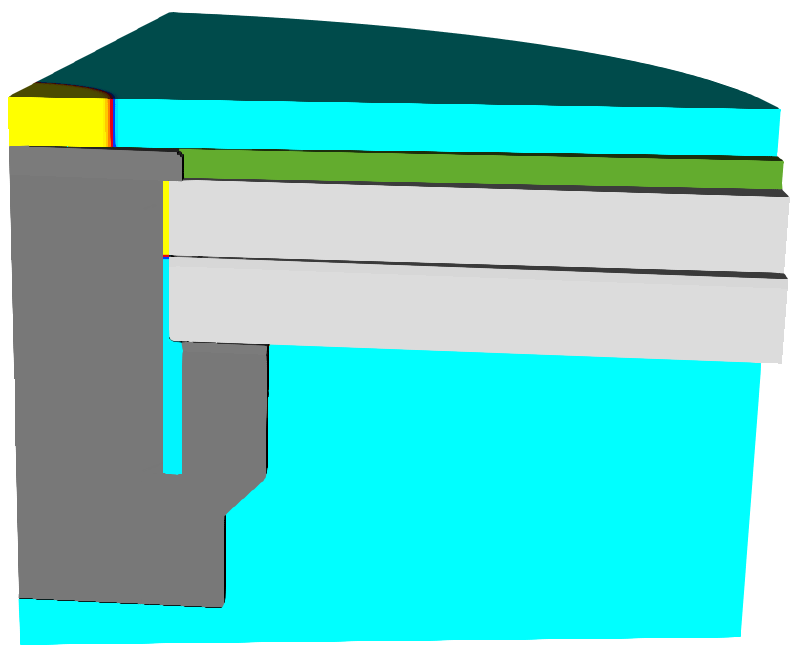}
   \caption{ An example fastener configuration with a pre-heated upper
     region of the fastener gap, visible as a yellow region between
     the fastener and the carbon composite panels.  All other features
     in this configuration are the same as in
     figure~\ref{fig:BoeingFastener_InitialAnnotated}. }
  \label{fig:BoeingFastener_PreHeating}
\end{figure}

The behaviour within the fastener gap, shown in
figures~\ref{fig:BoeingFastener_Snapshots} and
\ref{fig:BoeingFastener_Streamlines}, assumes that ionisation within
this region results only from mechanical effects at the gap interfaces
raising temperature and pressure.  However, electromagnetic effects
can lead to breakdown of the air within this region, and this offer an
alternative mechanism for the generation of plasma.  Modelling this
breakdown is a complex issue and is beyond the scope of this work, however,
such behaviour can be approximated by defining a pre-heated region within
the fastener gap. This approach is similar to the method used to initialise the
plasma arc from the electrode. The necessity to approximate the
breakdown of air in this gap is also reported by Kirchdoerfer et
al~\cite{kirchdoerfer2017}, who artificially augment the energy in the
gap between a fastener, panel and nut. In the work of Kirchdoerfer et
al~\cite{kirchdoerfer2017}, the energy in the confined gas void is
increased through the detonation of a small charge in the fastener gap
at a time corresponding to the moment when the local electric field
adjacent to the lower panel exceeds the dielectric strength of air,
3~MVm$^{-1}$. This raises the resulting pressure in the gap, and leads
to simulation results which better approximate experimental measurements. 
It is therefore of interest to establish a conductive path by
pre-heating a small section at the top of the gap which may then increase the
pressure in the remaining gap over the course of the
simulation. Understanding the effect of any approximation to
breakdown and the sensitivity this has on the pressure within the gap
is important because high pressure close to the bolt-nut interface is considered 
to result in energetic discharge (outgassing) from an interface of this type. An
example configuration with pre-heating within the fastener gap region
is shown in figure~\ref{fig:BoeingFastener_PreHeating}.

\begin{figure}[!ht]
  \centering
   \includegraphics[width=0.5\textwidth]{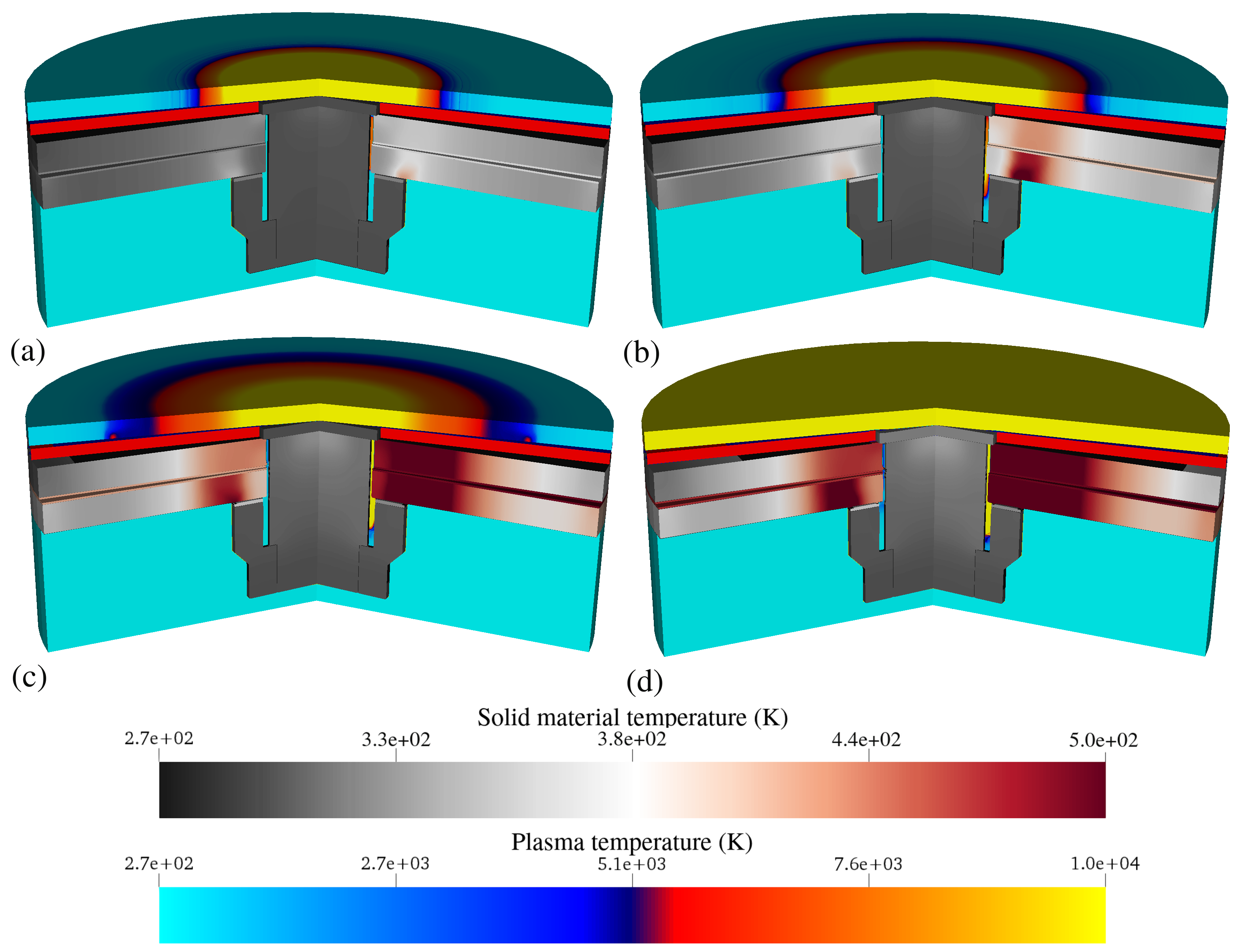}
   \caption{Comparison of the temperature between fasteners
     configurations with no pre-heating (left-half) and with
     pre-heating (right-half) of the upper region of the fastener
     gap. Snapshots are shown at times (a) 5~$\mu$s, (b) 10~$\mu$s,
     (c) 20~$\mu$s and (d) 40~$\mu$s.  It is clear that the pre-heated
     region leads to a large temperature rise within the gap, but also
     to greater temperatures within the panels.}
  \label{fig:BoeingFastener_Snapshots_BacktoBack}
\end{figure}

Figure~\ref{fig:BoeingFastener_Snapshots_BacktoBack} shows a
comparison of the temperature profile for configurations with and without a pre-heated gap. 
The temperature evolution for the fastener configuration with pre-heating is shown as
the right-half of each sub-figure, whilst the left-half is the case
without, with results reproduced from
figure~\ref{fig:BoeingFastener_Snapshots}.  It is clear that the high
temperature region in the upper section of the fastener gap, results
in significant evolution of the material in this gap, filling much of
it with plasma.  This evolution slows at later times as it expands
into the larger volume collar gap. Temperatures in this
region are sufficient for a partially ionised plasma to form, hence
electrical conductivity increases to provide a new current pathway
between the fastener and the surrounding carbon composite panels. The
current passage through the plasma in the gap consequently increases
the temperature and pressure.

In addition to the evolution within the fastener gap, there is also an
increase in temperature in the carbon composite panels for the
pre-heated case.  This is due to direct energy deposition in
the panels, since ionised material in the fastener gap provides a
highly conductive path for the majority of the current to
flow.

\begin{figure}[!ht]
  \centering
   \includegraphics[width=0.45\textwidth]{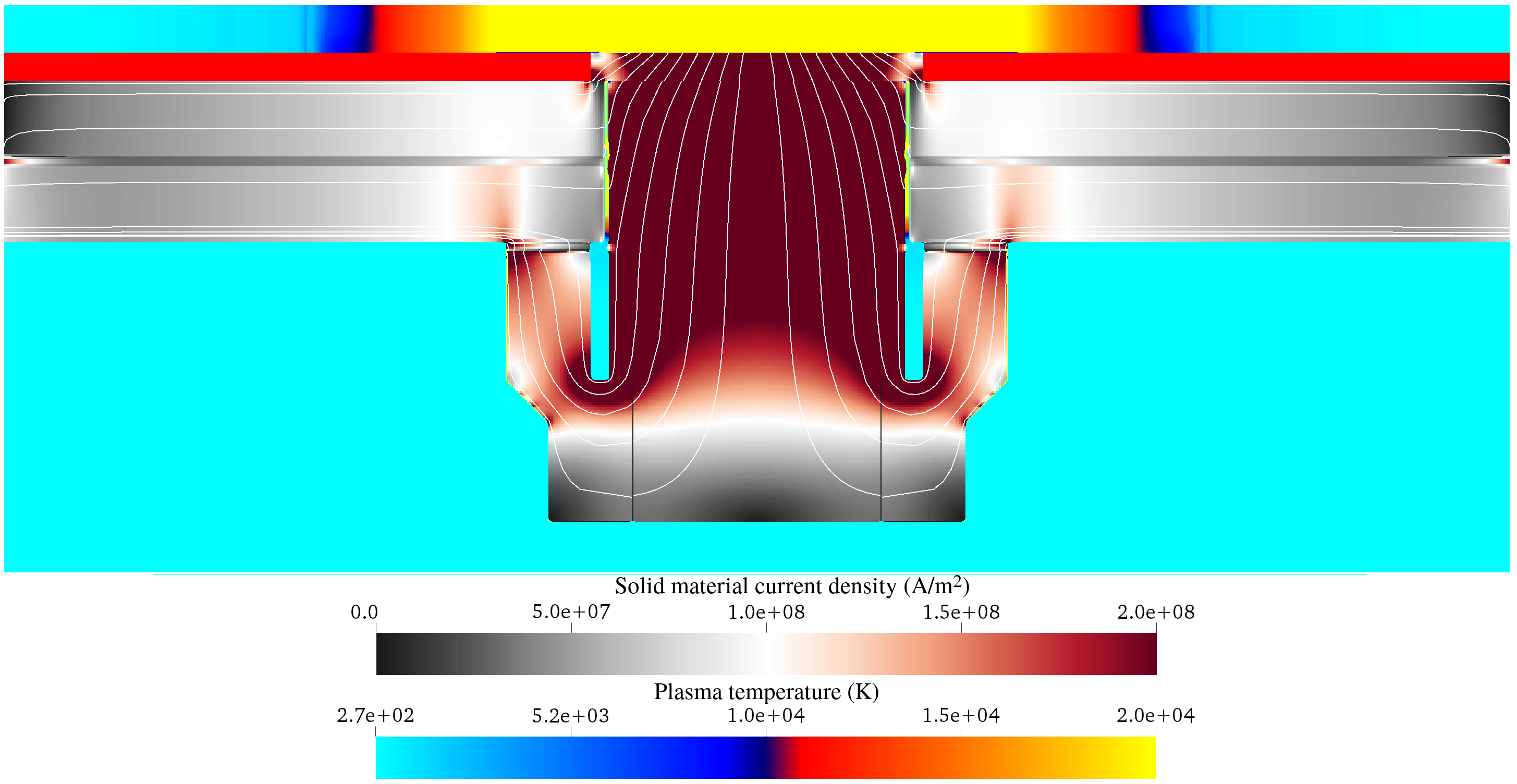}
   \caption{Current density magnitude and streamlines (white lines)
     within the substrate materials for the fastener configuration
     shown in figure~\ref{fig:BoeingFastener_Snapshots_BacktoBack} after
     10$\mu$s.  The pre-heated region within the fastener gap results
     in current flow throughout the panels, in comparison to
     figure~\ref{fig:BoeingFastener_Streamlines} where the current in
     confined to the top and bottom surface of the panels.}
  \label{fig:BoeingFastener_Streamlines_withPreheating}
\end{figure}

Figure~\ref{fig:BoeingFastener_Streamlines_withPreheating}
shows the current streamlines flowing through the panels.  
This can be compared with the case without pre-heating in figure~\ref{fig:BoeingFastener_Streamlines}, where current
flow is restricted to the top of the upper panel and the bottom of the
lower panel. The electrical conductivity in the plasma gap at this time
is still lower than the electrical conductivity in the titanium
fastener and shank, hence a significant proportion of the current
still passes through the fastener shank, nut and lower side of the
carbon composite panel, and directly between the fastener head and the
upper side of the carbon composite panel. The addition of a viable path
in figure~\ref{fig:BoeingFastener_Streamlines_withPreheating} across
the fastener gap further raises the plasma temperature, pressure and electrical
conductivity, reinforcing this route for current flow.
\\

\begin{figure}[!ht]
  \centering
   \includegraphics[width=0.38\textwidth]{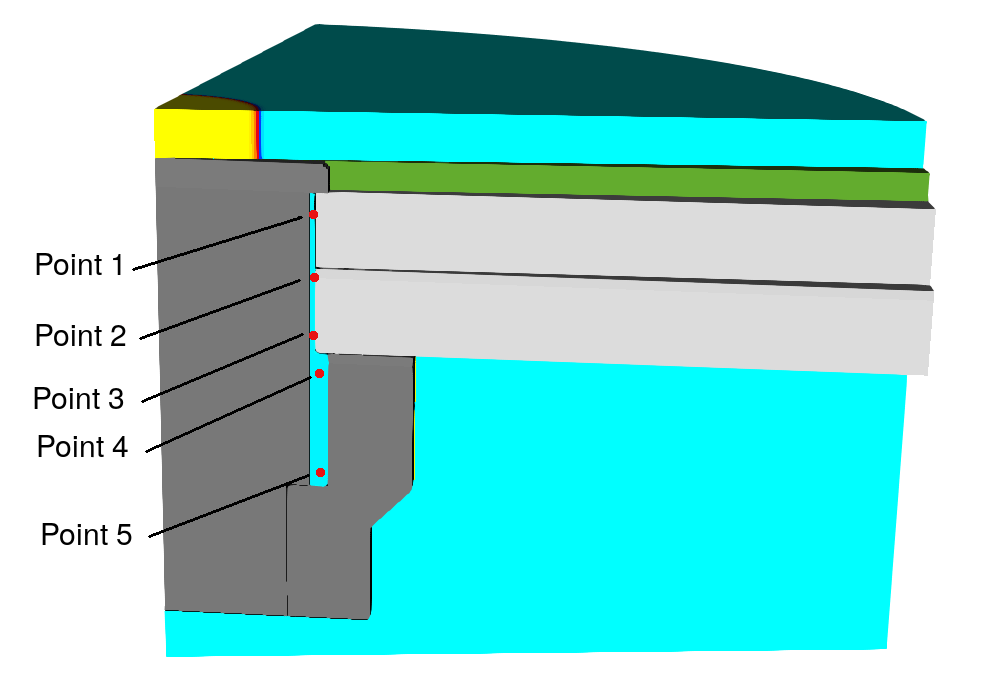}
   \caption{ Location of points for recording evolution data within
     the fastener gap}.
  \label{fig:BoeingFastener_DataPoints}
\end{figure}

\begin{figure}[!ht]
  \centering
   \includegraphics[width=0.45\textwidth]{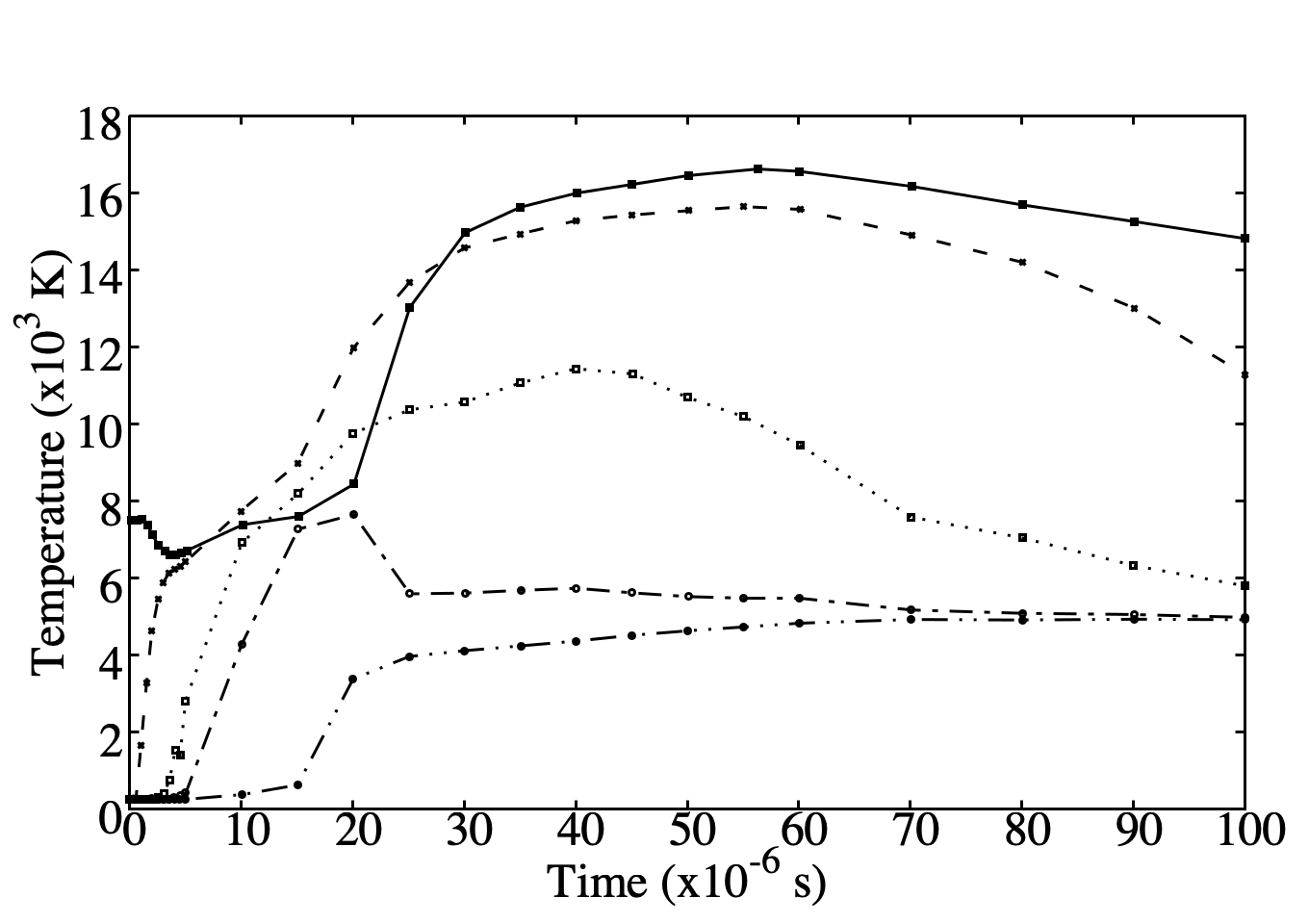}
   \caption{Temperature evolution at the five fixed data recording
     locations shown in figure~\ref{fig:BoeingFastener_DataPoints};
     solid line: point 1, dashed line and squares: point 2, dotted
     line: point 3, dashed line and circles: point 4, dotted and
     dashed line: point
     5. For points 1-3, the temperature rises at early times, but
     expansion into the wider region between the collar, and a
     reduction in current flow, leads to a drop in temperature at
     later times.  For the two points in the wider region of the gap,
     temperature initially rises, and then remains close to constant
     throughout the simulation.}
  \label{fig:BoeingFastener_TemperatureDevelopment}
\end{figure}

The behaviour of the material in the fastener gap can be monitored
over the course of the simulation by defining data collection
points. Five fixed spatial points are illustrated in figure~\ref{fig:BoeingFastener_DataPoints}. 
Three of these points monitor behaviour in the narrow section of the
fastener gap and two further data points are located in the wider cavity between
the fastener bolt and collar. The temperature over the first
100~$\mu$s is shown for each of these locations in
figure~\ref{fig:BoeingFastener_TemperatureDevelopment}.

The temperature at the highest location, point 1, in
figure~\ref{fig:BoeingFastener_TemperatureDevelopment} rises
gradually from the pre-heated temperature over the first
20~$\mu$s. This is then accompanied by a rapid rise in temperature
between 20~$\mu$s and 40~$\mu$s as this region becomes a viable
current path. The temperature continues to increase to 60~$\mu$s,
after which the reduction in input current and the expansion in to
the wider gap between the fastener and collar results in a drop in
temperature. Points 2-5, are initially located just outside the
pre-heated region, hence the material here is initially under ambient
conditions.  The initial discontinuity between the pre-heated and 
ambient regions results in a shock wave which travels along the
fastener gap, raising the pressure and temperature in the lower fastener gap region. 
This is visible through the initial rise in temperature at points 2 and 3, and also in the initial decrease in
temperature at point 1 over the first 5~$\mu$s. After this time, the current flow through the material is the dominant cause of
evolution, and the temperature continues to rise, with point 2
reaching a peak value of 1540~K at 55~$\mu$s. The subsequent decrease
in temperature at point 2 is greater than for point 1, falling to
1200~K at a time of 100~$\mu$s, whilst point 1 falls to 1420~K. Point
3 shows a similar temperature profile though the rise in temperature
from ambient conditions occurs later than at point
2. Figure~\ref{fig:BoeingFastener_TemperatureDevelopment} also
shows that the peak temperature at point 3 is again lower than for the
two higher locations, rising to 1145~K at 40~$\mu$s. The two points
inside the larger collar cavity, points 4 and 5, appear
to be outside the main current path, with the rise in temperature in
this region largely due to flow from the higher temperature regions
above. This continued movement from the upper cavity region to the
lower region causes the temperature at the lowest point, point 5, to
continue to increase over the entire course of the simulation, though
the final temperature here remains considerably lower than the upper
four points. This gradient highlights the difference in gap temperature (and hence
pressure) that can result from the existence of a viable current path
across only part of the fastener gap. In the next section, the data
collection points are used to investigate how the pressure in the gap
changes with alterations in the width of the narrow fastener gap.

\subsection{The effect of clearance gap size}
\label{sec:FastenerShankWidth}

One of the contributory factors in outgassing and thermal sparking
from fastener joints is considered to be the pressure rise in the
collar gap. Understanding the role of gap size on the pressure rise
close to the interface is therefore important for guiding design
choices. The present numerical approach can be used to investigate the
effects of geometry-related changes in pressure within the fastener
gap.

Using configurations similar to those presented in
section~\ref{sec:FastenerPreHeating}, the effect of changing the
radial distance between the fastener bolt and the carbon composite panel is
considered, for pre-heated and non pre-heated fastener gaps.
Seven cases are considered, with the distance between fastener and
panels referred to as the `shank gap'. The larger gap, between the
fastener and the collar, is termed the `collar gap' in this section and is
kept constant throughout. The shank gap is varied between 50~$\mu$m and
200~$\mu$m.  All other parameters, including the
larger collar gap and the panel thickness, are kept the same.

\begin{figure}[!ht]
  \centering
   \includegraphics[width=0.45\textwidth]{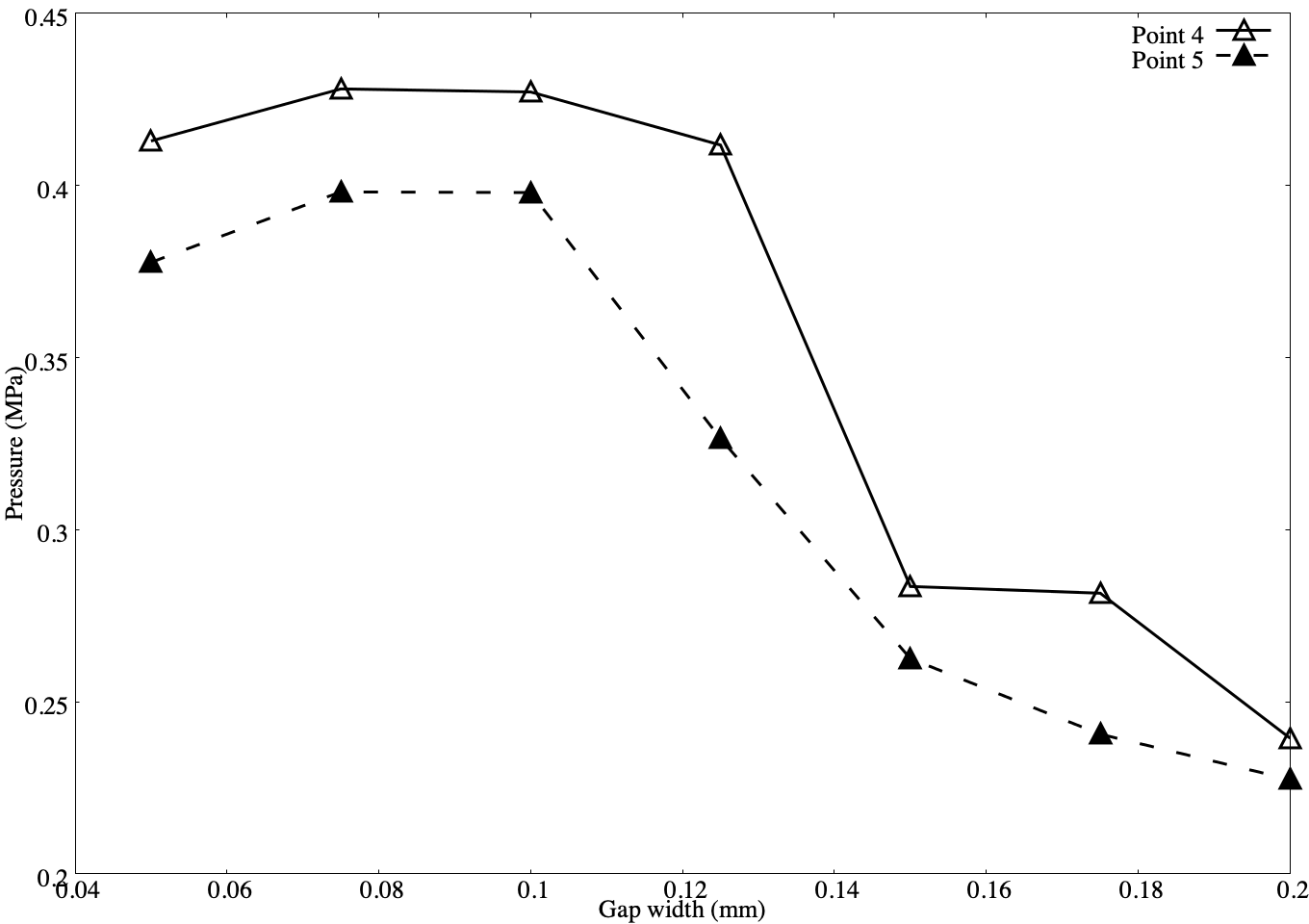}
   \caption{ The effects of the shank gap width on the pressure at
     data points 4 (hollow triangles) and 5 (filled triangles) after
     100~$\mu$s with no pre-heating within the fastener gap.  For
     narrow gaps, pressure at these points rises with increasing shank
     gap size, though as this gap gets wider still, there is then a
     drop in the pressure at these points. }
  \label{fig:BoeingFastener_GapWidth_NoPreHeating}
\end{figure}

The key region at which outgassing is likely to occur is where the collar meets
the carbon composite panel, hence the primary interest in this study
is to compare the pressure rise at data collection points 4 and 5, as identified in
figure~\ref{fig:BoeingFastener_DataPoints}. Figure~\ref{fig:BoeingFastener_GapWidth_NoPreHeating}
shows the pressure at these two points after 100~$\mu$s for the case
without pre-heating.  This time is chosen to
allow the pressure evolution to equilibrate between the collar and
shank gaps. The pressure is shown to vary significantly with the width
of the shank gap, though in all cases the pressure at point 4 (the higher of the
two locations) remains greater than at point 5. The
pressure decreases sharply at point 5 above a shank gap width of
0.1~mm, though it remains high at this width for point 4. This may
indicate a reduction in flow from the upper to the lower
region of the shank gap. The subsequent reduction in pressure at both 
data point locations for wider shank gaps highlights the decrease in
electrical conductivity at the top of the gap. It is also noted
that for all shank gap widths, the pressure is about two orders of
magnitude below those cited in experiments.

\begin{figure}[!ht]
  \centering
   \includegraphics[width=0.45\textwidth]{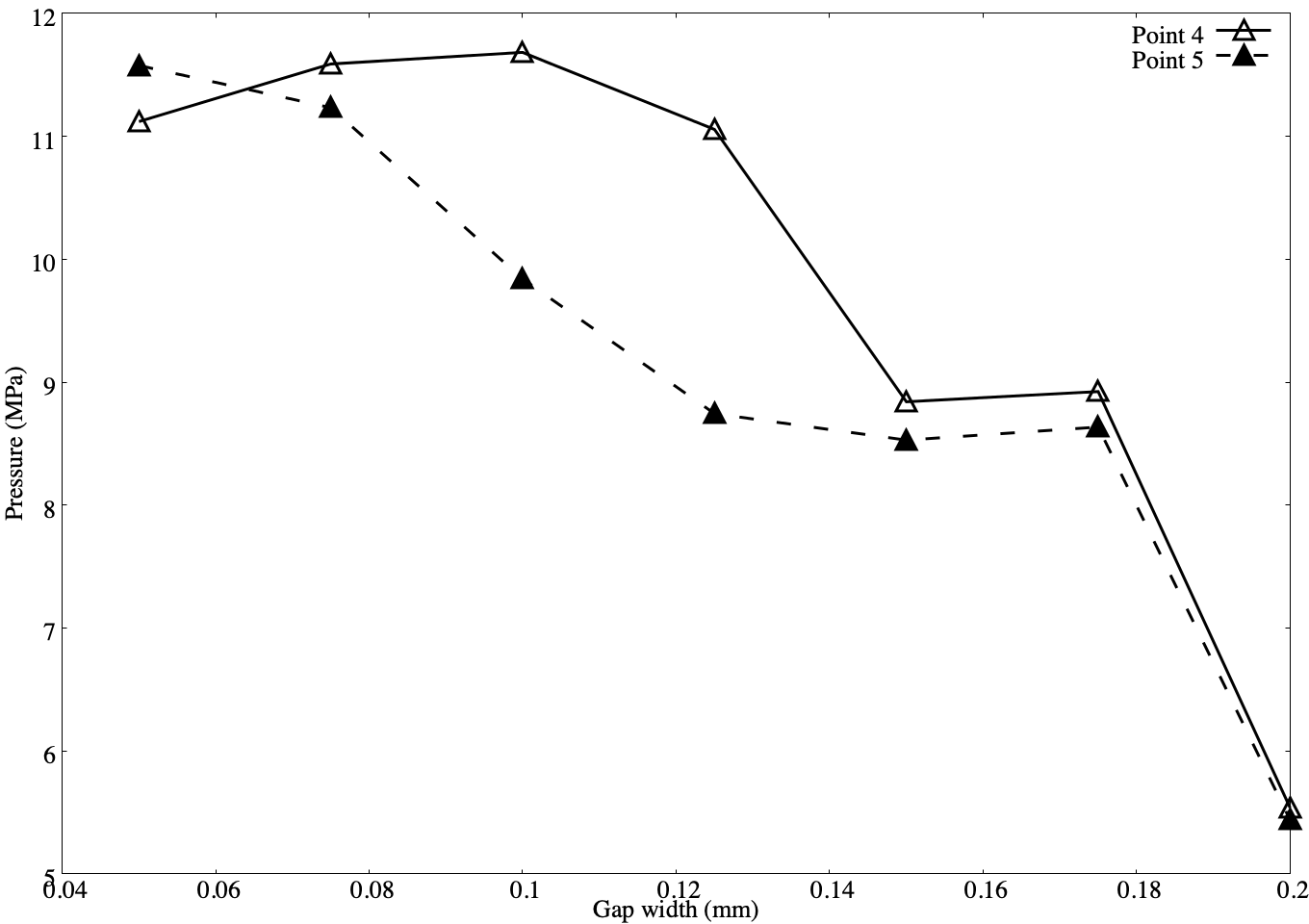}
   \caption{ The effects of shank gap width on the pressure at data
     points 4 (hollow triangles) and 5 (filled triangles) after
     100~$\mu$s with an initial pre-heated region at the top of the
     shank gap.  The effects of increasing the shank gap width are
     more pronounced for point 5, for which a wider gap correlates to
     a lower pressure.  For point 4, this behaviour is only seen for
     shank gaps above about 0.125~mm.}
  \label{fig:BoeingFastener_GapWidth_PreHeated}
\end{figure}

Figure~\ref{fig:BoeingFastener_GapWidth_PreHeated} shows the pressure
at points 4 and 5 after 100~$\mu$s for varying shank gap width where
the upper region of the shank gap has been pre-heated.  In all cases,
the pressures at both points are significantly greater than those
without pre-heating, in
figure~\ref{fig:BoeingFastener_GapWidth_NoPreHeating}.  In this case,
a maximum pressure of 11.8~MPa occurs at point 4 for the 0.1~mm shank
gap width. Interestingly, for the narrowest shank gap, 0.05~mm, the
pressure at point 5 is greater than that at point 4. This is not
mirrored in any of the other gap widths and may be associated with a
greater penetration of the high temperature fluid from the shank gap
into the collar gap. A similar trend to that shown in
figure~\ref{fig:BoeingFastener_GapWidth_NoPreHeating}, with pressure
decreasing as shank gap width increases is observed in
figure~\ref{fig:BoeingFastener_GapWidth_PreHeated}. Again, the drop in
pressure resulting from the increasing shank gap width at point 4 lags
behind that of point 5.  From analysis of the corresponding pressure
traces in the shank gap, this decrease in pressure appears to be
predominantly associated with a decrease in the length of time over
which the electrical conductivity in the upper regions of the shank
gap is sufficiently high for the fastener gap to remain a viable
current path.

The results in this section demonstrate that the multi-physics
methodology outlined in this paper is capable of computing the effect
of transient changes in current flow through a fastener geometry on
the pressure and temperature within the substrate materials, and
in gas-filled voids. These results also indicate that this
approach is capable of accounting for the influence of geometric
changes on the current distribution and associated thermal and
mechanical fastener behaviour. The increase in temperature and
pressure in the pre-heated gas results further highlights the 
influence of gas-filled gaps on the electrical and thermal 
development of the fastener as a whole.

The importance of considering direct lightning attachment in fastener
design is highlighted by the simulations presented in this paper. 
Understanding the path taken by the current through the fastener
geometry can lead to judicious use of dielectric layers to manipulate 
the current path. Potentially this can mean directing the current 
away from electrically sensitive components, minimising indirect 
attachment to remote components, or reducing the possibility of energetic discharge, thermal sparking and
edge glow. The simulations in this paper also provide confirmation of
existing best practice, that to minimise the possibility of
outgassing, the fastener design should maximise the electrical contact
between fastener components. This, in turn, minimises the contact
resistance and reduces the potential for ionisation of the gas-filled gaps
between components. Sections~\ref{sec:FastenerPreHeating}
and~\ref{sec:FastenerShankWidth} suggest that the plasma heating and
mass transport characteristics in regions of confined plasma can be
influenced by the size, shape and relative location of internal voids. 
The simulations presented in this work also suggest
that convective transport of the hot plasma can lead to a pressure
rise in connected void regions away from the direct current path.


\section{Conclusions}
\label{sec:Conclusions}

This paper presents a multi-physics methodology that provides dynamic,
non-linear coupling of a plasma arc with an elastoplastic
multi-material model description of an aerospace fastener
assembly. This methodology simultaneously solves hyperbolic partial
differential equations for each material to achieve a two-way coupled
system between the plasma arc and the fastener materials. The
advantage of this approach is that the transient changes in the
mechanical and electrodynamic properties for each material, and their
associated influence on the surrounding materials, are captured.

The ability for the numerical model to capture the dynamic influence
of changes in material choice, and layering design, on the electrical
current path and associated thermal and mechanical properties is
demonstrated. This highlights how a non-judicious use of dielectric
layers and the presence of unsealed internal gaps in the fastener
design may result in conditions such that structural and sparking
issues could occur under a transient, high current event. Joule
heating of the substrate materials in these regions can result in high
local stresses and material temperatures. Large potential differences
across internal gas-filled voids can cause ionisation and the
promotion of a high pressure plasma which is considered to be a
driving mechanism for energetic discharge from fasteners through
sparking. The inclusion of simple contact resistances in the numerical
model accounts for surface roughness, fibre ply pull-out and other
surface imperfections between fastener components. This could be
extended to include dynamically changing contact resistances to
account for transient changes in the mechanical and thermal loading of
components over the course of the lightning strike.

Additionally, the model presented in this work could be extended to
include a statistical approximation to the microscopic surface
imperfections that exist at the contact surface between adjacent
materials.  This would allow additional mechanisms for energetic
discharge to be studied, in particular thermal sparking, as detailed
by Odam et al.~\cite{odam1991lightning, odam2001factors}. 
Incorporating a fully anisotropic equation of state for
composite materials, for example the method of
Lukyanov~\cite{lukyanov2010equation}, would allow for
directionally-dependent effects to be studied.

The flexibility of the numerical model for undertaking parameter
studies is demonstrated with an example in which the dimensions of an
internal gap between fastener components is systematically
altered. This paper is intended to tentatively highlight the potential
for the multi-physics methodology to be used as an engineering tool in
the design and optimisation of aerospace components subject to plasma
arc attachment, or as a developmental aid in experiment design.


\section*{Acknowledgements}

The authors acknowledge the funding support of Boeing Research \&
Technology (BR\&T) through project number SSOW-BRT-L0516-0569. The
authors would also like to thank Micah Goldade, Philipp Boettcher and
Louisa Michael of BR\&T for technical input throughout this work.

\bibliographystyle{unsrt}
\bibliography{FastenerPaper_versionPostJAP.bib}

\end{document}